\documentclass[journal]{IEEEtran}

\usepackage{cuted}
\usepackage{multicol}
\usepackage{comment}
\usepackage{cite}
\usepackage{amsmath,amssymb,amsfonts,lipsum}
\usepackage{algorithmic}
\usepackage{graphicx}
\usepackage{textcomp}

\usepackage{cite}
\usepackage{amsmath,amssymb,amsfonts}
\usepackage{algorithmic}
\usepackage{graphicx}
\usepackage{epstopdf}
\usepackage{subfigure}
 \graphicspath{{\figures}}
\usepackage{textcomp}
\usepackage{xcolor}
\usepackage{verbatim}
\usepackage{makecell}
\usepackage{booktabs}
\usepackage{CJK}
\usepackage{multirow}
\usepackage{bbm}
\usepackage{amsfonts}
\usepackage{amsmath}
\usepackage{dsfont}

\usepackage{amsthm}
\usepackage{amssymb}

\usepackage{stfloats}
\usepackage{amsmath}
\usepackage{lipsum}

\usepackage{algorithm}
\usepackage{algorithmic}

\usepackage{multicol} % 
\makeatletter
\newcommand{\removelatexerror}{\let\@latex@error\@gobble}
\makeatother

\begin{document}

% \begin{CJK}{UTF8}{gbsn}
\newtheorem{theorem}{Theorem}
\newtheorem{lemma}{Lemma}

\title{
% Continual Learning for Edge Wireless Networks
Continual Reinforcement Learning for Digital Twin Synchronization Optimization

% Cost-efficient Sample Policy for Real-Time State Estimation  over Cellular Network
}
% \author{Haonan Tong, Ye Hu, Sihua Wang, Changchuan Yin\\
        % hntong@bupt.edu.cn; yeh17@vt.edu; sihuawang@bupt.edu.cn; ccyin@bupt.edu.cn}
       
% \author{Haonan~Tong,~\IEEEmembership{Member,~IEEE},~Zhaohui Yang,~\IEEEmembership{Member,~IEEE},~Sihua Wang,~\IEEEmembership{Member,~IEEE},\\
% ~and Changchuan Yin,~\IEEEmembership{Senior Member,~IEEE}
\vspace{0.cm}
\author{Haonan~Tong, 
\IEEEmembership{Student Member, IEEE},
Mingzhe~Chen, \IEEEmembership{Member, IEEE}, Jun Zhao, \IEEEmembership{Member, IEEE}, \\
Ye Hu, \IEEEmembership{Member, IEEE},  Zhaohui Yang, \IEEEmembership{Member, IEEE}, Yuchen Liu, \IEEEmembership{Member, IEEE}, \\
Changchuan~Yin, \IEEEmembership{Senior Member, IEEE}

% ~Sihua~Wang,~Zhaohui~Yang,
% ~Jun~Zhao, \textit{Member, IEEE}, \\
% ~Zhaohui~Yang, \textit{Member, IEEE}, \\
% Jun~Zhao, \textit{Member, IEEE} 
\thanks{

% This work was supported in part by Beijing Natural Science Foundation
% under Grant L223027, in part by the National Natural Science Foundation of
% China under Grants 61629101 and 61671086, 111 Project under Grant B17007,
% in part by Zhejiang Lab Program under Grant K2023QA0AL02, Zhejiang
% Science and Technology Program under Grant 2023C01021, BUPT Excellent
% Ph.D. Students Foundation under Grant CX2021114, and in part by China Scholarship Council.

H. Tong is with the Beijing Key Laboratory of Network System Architecture and Convergence,  the Beijing Advanced Information Network Laboratory, Beijing University of Posts and Telecommunications, Beijing, China, and also with the College of Computing and Data Science at Nanyang Technological University (NTU), Singapore.
Email: hntong@bupt.edu.cn. 
% N2308767J@e.ntu.edu.sg.

M. Chen is with the Department of Electrical and Computer Engineering
and Frost Institute for Data Science and Computing, University of Miami, Coral
Gables, FL, 33146, USA, Email: mingzhe.chen@miami.edu.

J. Zhao is with the 
College of Computing and Data Science 
% School of Computer Science and Engineering
at Nanyang Technological University (NTU), Singapore, Email: junzhao@ntu.edu.sg.

Y. Hu is with the Department of Industrial and Systems Engineering, University of Miami, Coral Gables, FL, 33146, USA, Email: yehu@miami.edu.

% Z. Yang is with the College of Information Science and Electronic Engineering, Zhejiang University, Hangzhou, Zhejiang 310027, China

Z. Yang is with the College of Information Science and Electronic Engineering, Zhejiang University, Hangzhou, Zhejiang 310027, China,
Email: yang\_zhaohui@zju.edu.cn.

Y. Liu is with the Department of Computer Science, North Carolina State
University, Raleigh, NC, 27695, USA, Email: yuchen.liu@ncsu.edu.

% Z. Yang is with the Zhejiang Lab, Hangzhou 311121, China, and also with the College of Information Science and Electronic Engineering, Zhejiang University, Hangzhou, Zhejiang 310027, China.
% % , and Zhejiang Provincial Key Lab of Information Processing, Communication and Networking (IPCAN), Hangzhou, Zhejiang, 310007, China. 
% Email: yang\_zhaohui@zju.edu.cn.

C. Yin is with the Beijing Key Laboratory of Network System Architecture and Convergence, and also with the Beijing Advanced Information Network Laboratory, Beijing University of Posts and Telecommunications, Beijing, 100876 China. Email: ccyin@bupt.edu.cn.
%Corresponding author: Changchuan Yin.
% A preliminary version of this work ~\cite{thn_conf_v_SCaware_DT} was accepted in the Proceedings of 2023 IEEE International Global Communications Conference (GLOBECOM).
% A conference version~\cite{thn_conf_v_SCaware_DT} of this paper has been published in Proceeding of IEEE Globecom 2023.
%The code for this paper is \text{https://github.com/wcsnSC/CRL\_MTR\_SAC}
% Z. Yang is with the Department of Electronic and Electrical Engineering, University College London, WC1E 6BT London, UK, Email: zhaohui.yang@ucl.ac.uk.
% Y. Hu and W. Saad are with the Wireless@VT, Bradley Department of Electrical and
% Computer Engineering, Virginia Tech, Blacksburg, VA, 24060, USA, Email: yeh17@vt.edu, walids@vt.edu.
}\vspace{-0.cm}   
}
\maketitle

\begin{abstract}
This article investigates the adaptive resource allocation scheme for digital twin~(DT) synchronization optimization over dynamic wireless networks.
% the mismatch minimization communication framework for remote state estimation in a digital twin~(DT) enabled smart factory with diverse wireless sensing devices.
In our considered model, a base station (BS) continuously collects factory physical object state data from wireless devices to build a real-time virtual DT system for factory event analysis.
% wireless sensing devices must continuously sample the factory physical object states and transmit sensing data to a base station, where the received sensing data is used to build a real-time DT mapping remotely to analyze and predict the object behaviors in the factory.
Due to continuous data transmission, maintaining DT synchronization must use extensive wireless resources.
To address this issue, a subset of devices is selected to transmit their sensing data, and resource block (RB) allocation is optimized.
% To address this issue, it is necessary to select a subset of devices to transmit their sensing data and optimize resource block (RB) allocation for devices. 
%to support the transmission of effective data.
% selectively collect effective data 
% that indicates a significant DT mapping mismatch.
% To address this issue, we reduce the amount of transmitted data by making the BS adaptively collect effective data that indicates a significant DT mapping mismatch.
%In particular, we first define the mismatch metric to measure the DT synchronization performance.
%Next, we 
This problem is formulated as a constrained Markov process (CMDP) problem that minimizes the long-term mismatch between the physical and virtual systems. % via solving the optimal device scheduling strategy over dynamic environments.
% scheduling a subset of devices at each time slot, that have effective data to be collected. 
To solve this CMDP, we first transform the problem into a dual problem that refines RB constraint impacts on device scheduling strategies. 
We then propose a continual reinforcement learning (CRL) algorithm  to solve the dual problem.
% multi-timescale replay for soft actor-critic~(MTR-SAC) based 
The CRL algorithm learns a stable policy across historical experiences for quick adaptation to dynamics in physical states and network capacity.
% time-varying physical states and changing resource budget.
% Then we propose  to learn a policy that adaptively schedules resource blocks~(RBs) to devices according to time-varying physical states and changing resource amount.
% in the long-term DT mapping.
% meet the transmission requests within the RB constraint, while considering packet error. 
Simulation results show that the CRL can adapt quickly to network capacity changes and reduce normalized root mean square error~(NRMSE) between physical and virtual states by up to 55.2\%, using the same RB number as traditional methods.

% the number of consumed RBs by up to 55.2\% compared to traditional scheduling methods, with the same mapping accuracy. 

\end{abstract}

\section{Introduction}
Future wireless networks must support interactive applications such as digital twins (DTs)~\cite{DTN_survey} and Metaverse~\cite{Metaverse}. 
These applications necessitate the mapping of a physical system into a synchronized virtual system, facilitated by ubiquitous and continuous sensing data transmission~\cite{DTN_network, TMC_DT_Service_Provisioning}. 
However, achieving continuous synchronization between a physical system and its corresponding virtual counterpart faces several challenges.
% These applications require to map a physical system into a synchronized virtual system
% with ubiquitous and continuous sensing data transmission~\cite{DTN_network}.
% However, maintaining a continuous synchronization between a physical system and its corresponding virtual system faces several challenges.
First, the mapping from the physical system to a virtual system requires extensive sensing data transmission, which may be hard to achieve in a wireless network with constrained communication resources~\cite{Comm_eff_DT, SLA_Quality_Assurance}.
% thus achieving accurate physical network state estimation with limited communication resources is challenging~\cite{Comm_eff_DT}.
Second, since the data required for the mapping is transmitted over wireless channels, the dynamics (i.e., channel conditions) of wireless environment will significantly affect synchronization between the virtual system  and physical system~\cite{syn_IoV, DT_syn_XMShen}.

% The strategy to mapping optimization must evolve along with the physical system.

% dynamic wireless environment changes the communication resource budget thus challenging the adaptivity of resource allocation strategy~\cite{syn_IoV,DT_syn_XMShen}.

% dynamic network factors such as time-varying physical states and wireless environment fluctuations will challenge the resource allocation scheme to maintain mapping synchronization

% Hence, efficient remote state estimation and adaptive wireless resource allocation mechanism are vital for wireless networks supporting prolonged virtual mapping in emerging applications. 

% Therefore, achieving efficient state estimation is a vital technology for future wireless networks. 
% Semantic communication, which aims to transmit the meaning of messages beyond the symbol level through semantic extraction, could be a powerful enabler of efficient state estimation~\cite{Semcom_survey}.

% DT是资源受限的, 这些工作都证明感知精度重要，DT对于 efficient remote estimation要求很高
% remote estimation 是咋演进的，但是没有考虑长期演进的事情，
% 考虑长期演进的事情，没有考虑实际在线持续性的学习与算法部署，

% 现有工作指明资源是重要的，但是没有具体分析有限资源对于感知的影响，有些机制研究了单一终端的采样机制，但是没有考虑多终端对资源的竞争关系，尽管现有的多终端资源分配进展到了强化学习的方式，但是还没有能适应动态环境的能力。

\vspace{-0.3cm}
\subsection{Related Works}

Recently, several works such as~\cite{zhuang_DT_planning_reconf,TMC_DT_FL,xyt_GC,adaptive_FL_DT,DYY_Adapt_DT, DT_enhenced,chixu,zehuixiong} have studied the applications of DT for wireless networks with limited resources.
% Prior research on DT in wireless network has studied wide range of issues caused by the limited wireless resources~\cite{zhuang_DT_planning_reconf, xyt_JSAC, 
The work in~\cite{zhuang_DT_planning_reconf} developed DT empowered network framework to optimize computational power, storage resource, and communication resource allocation.
The authors in~\cite{TMC_DT_FL,xyt_GC,adaptive_FL_DT} used DT to capture the computational capability of each device to improve federated learning performance under communication and computation resource constraints.
In~\cite{DYY_Adapt_DT}, the authors used DT to predict task requirements and network topology so as to reduce task offloading latency and network congestion. The work in~\cite{DT_enhenced} used DT to simulate the dynamic bandwidth budget, packet arrival, and throughputs of a network and use this information to improve the spectrum efficiency.
{\color{black} The work in~\cite{chixu} utilized DT to simulate the competition for communication and computing resources among multiple devices, where device state estimation errors were corrected by DT, thereby enhancing the performance of the network resource management strategy.
Furthermore, in~\cite{zehuixiong}, DT was introduced to capture the high mobility of aerial communication nodes, enabling network behavior analysis in non-terrestrial networks.}
However, most of these works in~\cite{zhuang_DT_planning_reconf,TMC_DT_FL, xyt_GC, adaptive_FL_DT, DYY_Adapt_DT,DT_enhenced,chixu,zehuixiong} that focused on DT applications did not consider how limited wireless resources (i.e., transmit power and spectrum) affect the DT synchronization, which requires continuous data transmission.

The works in~\cite{JCC_Fwk_ZHY,Iot_assist_DT_syn,DT_syn_ZHY, UAV_DY_syn,Data_syn_IoV} have studied efficient communication mechanisms to optimize DT synchronization. 
The work in~\cite{JCC_Fwk_ZHY} developed a joint communication and computation framework for DT synchronization to reduce communication latency under DT model accuracy constraints. 
In~\cite{Iot_assist_DT_syn}, the authors proposed a hierarchical data collection framework in multi-layer networks to optimize DT synchronization with limited spectrum resources.
The work in~\cite{DT_syn_ZHY} analyzed the impacts of communication delay on the DT model accuracy. 
In~\cite{UAV_DY_syn}, the authors adjusted the device schedule, transmit power, and computing frequency to optimize DT synchronization in unmanned aerial vehicle (UAV)-assisted edge computing networks.
The authors in~\cite{Data_syn_IoV} proposed a game theory based network access algorithm to efficiently synchronize the data between the DT and the vehicle.
However, most of the works in~\cite{JCC_Fwk_ZHY,Iot_assist_DT_syn,DT_syn_ZHY, UAV_DY_syn,Data_syn_IoV} assumed that physical objects change according to predetermined probabilities.
% thus ignoring the adaptability of communication mechanisms.
% the physical object changes in a mode modeled by predetermined functions.
Due to the dynamic nature of the physical system, the sensing data for building DT actually changes with unpredictable randomness, which is difficult to be modeled.
Hence, the sensing data transmission methods for DT synchronization must be adaptive to the changes of the physical system dynamics.
% Therefore, the algorithms designed in \cite{JCC_Fwk_ZHY,Iot_assist_DT_syn,DT_syn_ZHY, UAV_DY_syn,Data_syn_IoV} cannot adapt to dynamic wireless networks where the sensing data randomly changes over time.  

% according to the designed models.

% where the prior information cannot be obtained in advance thus  
% the random changes of sensing data for building DT cannot be accurately modeled without prior information.

%maintain the DT synchronization when the environment changes during the continuous DT mapping.
\vspace{-0.cm}
{\color{black}
% DT的资源分配跟无线资源管理有紧密的关系.
The works in~\cite{revirewer3_3,reviewer3_1,reviewer3_2,RL_demandvary_VFN_mig, Lu_adaptive_access,Cjy_DT, UAV_DT} have studied adaptive network resource management methods to guarantee data transmission performance in dynamic environments.
In particular, the works in~\cite{revirewer3_3,reviewer3_1,reviewer3_2} analyzed machine learning based resource management in dynamic wireless network situations.
The work in~\cite{revirewer3_3} analyzed radio resource scheduling problem with diverse transmission requirements as a Markov decision process~(MDP) and demonstrated that reinforcement learning~(RL) was an effective solution to the complex problem with dynamics.
The authors in~\cite{reviewer3_1} proposed RL based dynamic resource allocation method to accommodate high mobility in 5G networks while satisfying heterogeneous user requirements and network architectures.
In~\cite{reviewer3_2}, the authors improved network efficiency via an RL based dynamic power control scheme and verified RL can achieve low protocol overheads for practical applications.}
Furthermore, the authors in~\cite{RL_demandvary_VFN_mig, Lu_adaptive_access,Cjy_DT, UAV_DT} have studied the use of machine learning for DT optimization in dynamic wireless networks.
% To adapt to the dynamics of wireless network, the prior arts~\cite{RL_demandvary_VFN_mig, Lu_adaptive_access,Cjy_DT, UAV_DT} have studied the use of machine learning to optimize DT network performance. 
The work in~\cite{RL_demandvary_VFN_mig} used RL to minimize the energy consumption of network virtual function~(NFV) migrations in a DT network with dynamic NFV communication and computation resource requirements. 
% The authors in~\cite{Dist_RL_Min_AoI} conducted a distributed RL based sensor sampling scheme to learn the dynamics of sensed physical processes and spectrum competitions among multiple sensors, to guarantee the freshness of sensing data.
The authors in~\cite{Lu_adaptive_access} used RL to find the optimal target server for DT migration with varying network topologies.
The work in~\cite{Cjy_DT} proposed a multi-agent RL method to learn the change of vehicular topology and resource requirements of computation tasks, so as to minimize the task offloading costs in DT vehicular network.
The authors in~\cite{UAV_DT} used RL to learn the dynamic UAV connection topology and computation task requirements, thereby maximizing the computing resource efficiency by adaptive task assignment in UAV-assisted DT networks.
However, the RL methods designed in~\cite{RL_demandvary_VFN_mig,  Lu_adaptive_access, Cjy_DT, UAV_DT} cannot adapt to an unseen %changed 
environment
% cannot continually optimize DT synchronization, 
%for they are often overfitting to the given training tasks.
% for the continual DT mapping can not 
% adapt fast to the newly changed wireless environments, which are not suitable for continually optimizing DT synchronization 
since the hyper-parameters and exploration strategies of the RL methods in~\cite{RL_demandvary_VFN_mig,  Lu_adaptive_access, Cjy_DT, UAV_DT} are specifically designed to fit the training tasks in a given environment. 
Once the RL agent encounters an unseen environment, the specifically trained RL methods may not be able to find an optimal solution.
% In addition, the RL that learned dynamic service requirements cannot adapt to the changed resource budget which changed the constraint of the task, rather than just transfer the state.
Since DT requires a continuous mapping between the physical and virtual systems, the designed DT optimization methods must adapt to the unseen environment, i.e., dynamic wireless network capacity and unseen physical system status.

\subsection{Contribution}

The main contribution of this work is a novel DT framework that enables a base station (BS) to continually optimize DT synchronization {\color{black} via communication resource scheduling} in a dynamic wireless network where the status of physical system changes over time.
% environments with limited communication resources.
%\textit{To the best of our knowledge, this is the first work to adaptively maintain DT synchronization in dynamic environments using continual reinforcement learning solution.
%}   
The contributions of this article include

% To this end, we are motivated to propose a machine learning based adaptive resource allocation policy to perceive the significant mismatch of the state data and adaptively adjust sample strategy according to time-varying physical states and network capacity.  
% The contributions of this article mainly include

% \color{orange}

\begin{itemize}
 
\item 
We develop a digital network twin framework, where a BS collects physical object state data from wireless devices for the generation of a virtual system. %synchronized DT mapping.
Due to the limited wireless resource blocks~(RBs) for DT mapping, only a subset of devices can transmit their data to the BS for virtual system update.
Hence, the BS must select a subset of devices and optimize RB allocation for these devices to improve the synchronization between the physical and virtual systems.

% wireless devices only sample and send state data on demand with the indicator of mismatch-aware metric and the wireless resources are allocated centrally by DT to maintain the DT mapping synchronization. 

\item  %We define a mismatch metric to evaluate the DT synchronization performance.
%Then
We formulate this device scheduling and RB allocation problem as a constrained Markov process~(CMDP) problem whose goal is to minimize the long-term mismatch between the physical system and the virtual system.
% The problem optimization variables include device selection and RB allocation. 
% The problem is optimized by solving the device scheduling .
% optimization variable is the device scheduling 
The packet errors, as well as the dynamics in terms of physical object states and wireless network capacity, are considered in the problem.
% {\color{black}and ... }
%according to the time-varying physical states and changing RB budget. 

\item  To solve the problem, we first transform the problem into a dual problem that refines the impacts of RB constraints on device scheduling strategies.
Then, we propose a continual reinforcement learning~(CRL) algorithm to find the optimal device scheduling strategies.
In particular, a multi-timescale replay~(MTR) buffer is first introduced in the CRL algorithm for extracting common knowledge of strategies from both old and fresh experiences.
Next, a modified soft actor critic~(SAC) agent is used to enhance the exploration of the scheduling strategies while satisfying RB consumption  constraints. 
With this mechanism, the proposed CRL algorithm can quickly converge to an effective strategy for an unseen physical system status.
\end{itemize}
Simulations with the Intel Berkeley research lab sensor data and user trajectory locations 
%have been conducted. 
%The simulation results 
show that, compared to traditional methods that schedule device at fixed interval time slots, the proposed CRL based device scheduling algorithm can reduce the normalized root mean square error~(NRMSE) of the estimated virtual states by up to 55.2\%.
% \textit{To the best of our knowledge, this is the first work that perform continual DT synchronization over dynamic wireless networks.}
%nd can quickly adapt the device scheduling policy to the changing RB number.

The remainder of this paper is organized as follows. 
The system model and the problem formulation are illustrated in Section II. 
The proposed CRL algorithm for device scheduling in continual DT synchronization optimization is introduced in Section III.
In Section IV, simulation results are presented and discussed.
Finally, the conclusions are drawn in Section V.

\begin{figure*}[t]
    \centering
    \includegraphics[width=0.9\textwidth, height =0.36\textwidth%,bb= 50 50 520 480
    ]{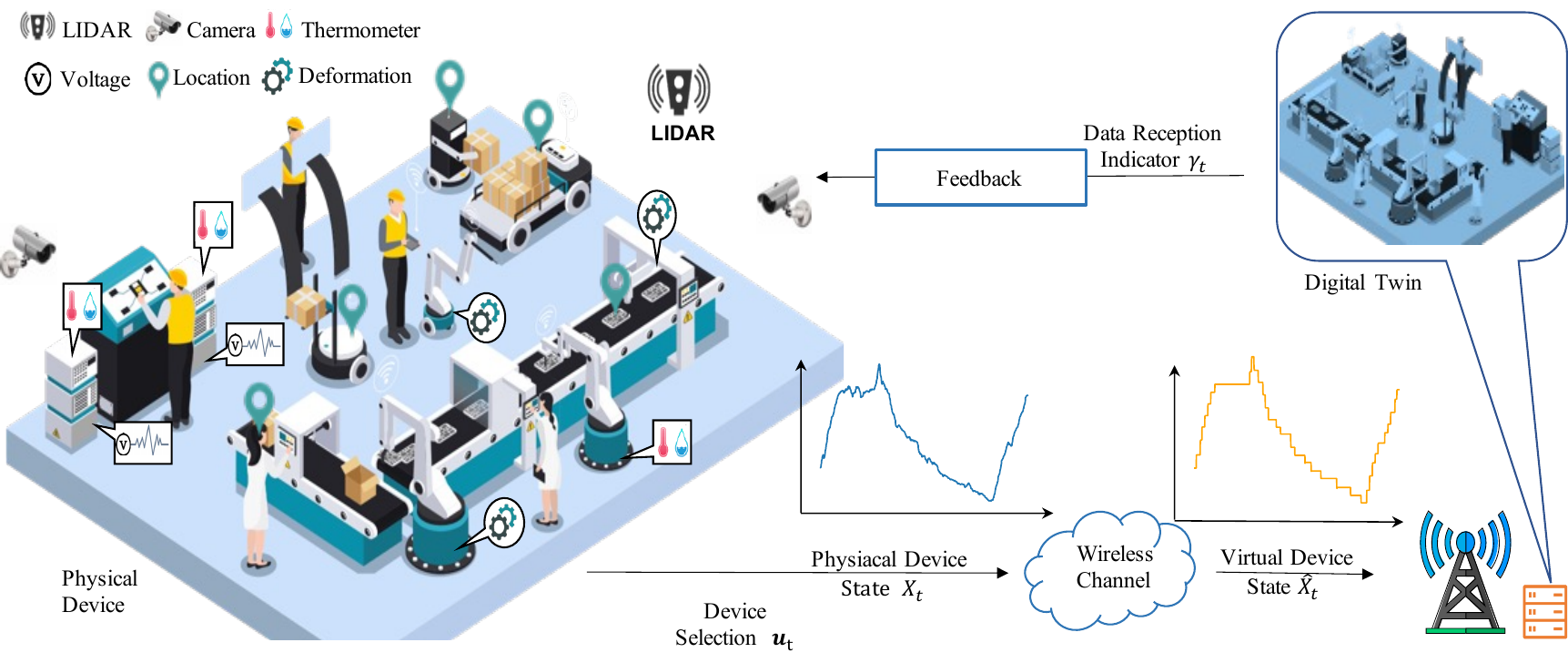}
    \setlength{\abovecaptionskip}{0.cm}
    \caption{DT enabled smart factory architecture. }\label{fig:mapping}
    \vspace{-0.cm}
\end{figure*}

% 基站学习一切，
% 所有的终端建立一个统一的误差可以评估多种相似度，然后 BS pull 传输需求

% 为了减少数据传输量，BS要识别误差，不调度的情况下，Device 不进行传输。

% Baseline 要改，改成随机调度策略。

\section{System Model and Problem Formulation}
% \color{orange}
% 具体明确一下，触发发送的机制，最后一项应该也要加阈值进来。

% 考虑一下，几个阈值函数是不是要放在后面Section IV 介绍。
% \color{black}

Consider a DT enabled smart factory {\color{black}in 5G wireless network} as shown in Fig.~\ref{fig:mapping}, in which a set $\mathcal{N}$ of $N$ devices (i.e. thermo-hygrometers, positioning sensors, monitoring cameras, and LiDARs) monitor factory from different perspectives i.e., room temperatures, robot motions, {workstation} status, %monitoring videos,
and three-dimensional factory layout. %models. 
{\color{black}Equipped with edge servers, a BS integrates communication and computational resources to collect sensing data from these devices for DT creation and synchronization~\cite{3GPP_R15_sys}.}
% {\color{black}Equipped with a server, a BS }collects the sensing data from these devices for DT creation and synchronization.
The DT will keep mapping the sensed physical object states into real-time virtual object states, to support continuous simulation and prediction of object behaviors. 
Notice that, DT mapping mismatch, the difference between the physical states and virtual states, inevitably exists because of the transmission delay and errors. 
Meanwhile, due to the limited wireless resources, %at the BS, it is not feasible for the BS to continuously collect sensing data from all the devices in real time. In such context, 
the BS can only selectively schedule a subset of the devices to collect effective sensing data that can revise the DT mapping mismatch.

% must implement efficient communication scheme by scheduling a subset of the devices, and allocate the scheduled devices resource blocks (RBs) to collect effective sensing data that revises the DT mapping mismatch.
% for the sensing data transmission. 
% Meanwhile, the BS must implement efficient communication scheme by collecting effective sensing data that revises the DT mapping mismatch. 
% For instance, the data of thermometer is  collected only when significant temperature changes occur.

% the cameras only upload videos when there are significant changes of the objects in the view. 
% LiDARs only upload 3D models when devices have significant changes, etc.

In the considered system, 
% once the  devices are scheduled, the devices transmit the sensing data through wireless channels to the BS.
once the devices are scheduled, they transmit sensing data to the BS via wireless channels.
% all devices transmit sensing data 
However, due to the inherent uncertainty of wireless channels, packet errors may occur during the sensing data transmission.
Hence, in the considered system, the BS can only update the virtual states at DT when the received sensing data does not have errors.
After receiving a packet, the BS responds with a feedback message (ACK/NACK).
% , which includes a data reception indicator as shown in Fig.~\ref{fig:mapping}.
Given the feedback message, the device can obtain the latest virtual state at DT.
Only after ACK is fed back, the device is informed that the virtual state has been updated into the most recent transmitted sensing data; otherwise, the virtual state stays.
In this way, the device can calculate the DT mapping mismatch by comparing the current observed physical state with the virtual state.
When scheduling a device to collect the sensing data, the BS can also collect the DT mismatch information of the device, which can be used to adjust the device schedule policy.
% With the feedback message, the device calculates the  DT mapping mismatch  by comparing the current observed state with the  virtual device state informed by the DT, and then can request RBs from the BS for sensing data transmission.
Next, we first introduce the mismatch model to measure the synchronization performance of the DT system.
% timeless of sensing data that indicates events.
Then, we formulate a mismatch minimization problem.
Table~\ref{tab:notation} provides a summary of the notations in this paper.
\color{black}

\begin{table}[t]
  \centering
  \renewcommand{\arraystretch}{1.5}
  \caption{LIST OF NOTATIONS}
  \label{tab:notation}
  \footnotesize
  \begin{tabular}{|c||c|}
    \hline
    \textbf{Notation} & \textbf{Description} \\
    \hline
    $N$ & Number of devices \\
    \hline
    $X_{n,t}$ & Physical state of device $n$ at time slot $t$ \\
    \hline
    $\widehat{X}_{n,t}(u_{n,t})$ & Virtual state of sensor $n$ at time slot $t$ \\
    \hline
    $u_{n,t}$ & Device selection index for device $n$ at time slot $t$ \\
    \hline
    $r_{n, t}(u_{n,t})$ & Data rate of device $n$ at time slot $t$ \\
    \hline
    $b_{n}$ & Number of occupied RBs by device $n$ \\
    \hline
    $\xi_{n}^{\mathrm{th}}$ & Threshold of mismatch function for device $n$ \\ 
    \hline
    $W$ & Bandwidth of each RB \\
    \hline
    $P_{n}^{\mathrm{u}}$ & Transmit power of device $n$ \\
    \hline
    $\sigma_{N}^{2}$ & Variance of AWGN \\
    \hline
    $o_{n}$ & Rayleigh fading factor of device $n$ \\
    \hline
    $d_{n}$ & Distance between device $n$ to the BS\\
    \hline
    $h_{n, t}$ & Channel gain of device $n$ \\
    \hline
    $L_{n}$ & Data size of the sensing data from device $n$ \\
    \hline
    $p_{n}$ & Packet error rate of the transmission of device $n$ \\
    \hline
    $m$ & Waterfall threshold in packet error function \\

    \hline
    $D_{n,t}(u_{n,t})$ & Uplink transmission delay for $X_{n,t}$ \\
    \hline
    $\gamma_{n,t}$ & Data reception indicator \\
    \hline
    $Z_{n,t}(u_{n,t})$ & Mismatch of device $n$ at time slot $t$ \\
    \hline
    $\boldsymbol{u}_t \in  \mathbb{R}^{1 \times N}$ & Transmission decision vector of devices \\
    \hline
    $\boldsymbol{b}_t \in  \mathbb{R}^{1 \times N}$ & RB occupation number vector \\
    \hline
    $M$ & Number of RBs at the BS for DT \\
    \hline
  \end{tabular}
\end{table}

% \color{orange}
% 我们首先决定了什么时候 device 是发送数据的，通过感知信源信宿的失配值，我们选择了误差显著的有效的数据，用最少得数据传输量来校正失配数据。

% 其次关注了校正失配数据传输的时效性，在考虑丢包的情况下。

% 首先，我们确定了失配的函数，由 xxx 构成。这些函数是阈值， 有 X  有X'

% 然后，DT的同步性能由这些失配数据传输的时效性来决定。因此，我们考虑了AoII

% AoII 是由BS的RB分配来决定的，因此将这些动态建模为一个MDP序列，

% 最终，问题的建模为长期AoII 最小化问题，
% \color{black}

\subsection{Mismatch Model}
Here we introduce the mismatch model, {\color{black} which is  extracted from both physical and virtual states to measure DT synchronization}.
At discrete transmission time slot $t$, the BS schedules device $n$ to upload its sampled sensing data $X_{n,t}$ using orthogonal frequency division multiple access (OFDMA). 
The BS updates the virtual device state $\widehat{X}_{n,t}$ at DT with the sensing data $X_{n,t}$. 
% Due to the limited wireless resource, we assume that device $n$ only transmits the sensing data for reducing the mismatch between $X_{n,t}$ and $\widehat{X}_{n,t}$, rather than timely delivering the packets of the sensed stochastic processes.
% Hence, we first decide which time slot for the devices to send sensing data.
% Given the data sending scheme, we define the  mismatch model, which indicates the time span since the DT mapping mismatch occurs.
We assume that the BS can allocate a set $\mathcal{M}$ of $M$ uplink orthogonal RBs to the scheduled devices and each RB can be allocated to at most one device.
The data rate of the sensing data uplink transmission from device $n$ to the BS at discrete time slot $t$ is given by (in bits/s)
\begin{equation} \label{equ:shannon}
r_{n, t}(u_{n,t})= u_{n,t} b_{n} \color{black} W \log _{2}\left(1+\frac{P_{n}^{\mathrm{u}} h_{n, t}}{N_0 b_n  W }\right),
\end{equation}
where $u_{n,t}\in\{0,1\}$ is a device schedule index  with $u_{n,t}=1$ indicating that device $n$ is selected to transmit the sensing data $X_{n,t}$ at time slot $t$, and $u_{n,t}=0$, otherwise;
$b_{n}$ is the number of RBs requested by device $n$ from the BS; $W$ is the bandwidth of each RB; $P_{n}^{\mathrm{u}}$ is the transmit power of device $n$; $h_{n, t}=o_n d_n^{-2}$ is the channel gain with $o_n$ being the Rayleigh fading factor and $d_n$ being the distance between device $n$ and the BS, and $N_0$ is the noise
power spectral density.
% $\sigma_{N}^{2}$ is the variance of  additive white Gaussian noise~(AWGN).
% Since the server has adequate communication and computation resources, we mainly consider the resource constraints of the uplinks from the devices and the server.

Let $L_{n}$ be the data size of the sensing data of device $n$, the uplink transmission delay of device $n$ transmitting $X_{n,t}$ to the BS is
\begin{equation}
D_{n,t} (u_{n,t}) = \frac{L_{n}}{r_{n, t}(u_{n,t})}.
\end{equation}
The packet error rate is defined as 
\begin{equation}
    p_{n}=\mathbb{E}_{h_n,t} \left(1-\exp\left(-\frac{m N_0 b_n W}{P_n^{\mathrm{u}} h_{n,t}}\right)\right), 
\end{equation}
where $\mathbb{E}_{h_{n,t}}\left( \cdot \right)$ is the expectation taken over the channel gain $h_{n,t}$  and  $m$ is  a waterfall threshold~\cite{pack_e}.
Then, the probability that the BS correctly receives  a sensing data packet uploaded by device $n$ is $1-p_n$. 
Hence, the data reception indicator $\gamma_{n,t}$ is given by 
\begin{equation} \label{equ:Bernoulli}
    \gamma_{n,t} \sim B\left(1- p_n\right),
\end{equation}
where $B(\cdot)$ is the Bernoulli distribution function, $\gamma_{n,t} =1 $ indicates the sensing data $X_{n,t}$ of device $n$ is received without errors at time slot $t$, and $\gamma_{n,t} =0 $, otherwise.

% We define the probability that the BS correctly receives a sensing data packet transmitted by device $n$ is $1-p_n$, where $p_{n}=\mathbb{E}_{h_n,t} \left(1-\exp\left(-\frac{m\sigma_N^2}{P_n^{\mathrm{u}} h_{n,t}}\right)\right)$ is the packet error rate with $m$ being a waterfall threshold~\cite{pack_e}.
% Hence, $\gamma_{n,t}$ obeys a 
% Bernoulli distribution and is expressed as
% \begin{equation} \label{equ:Bernoulli}
%     \gamma_{n,t} \sim B\left(1- p_n\right).
% \end{equation}

Upon the reception of sensing data, the virtual state at DT $\widehat{X}_{n,t}$ is updated with the latest received data, otherwise $\widehat{X}_{n,t}$ remains the same as the last time slot, as shown in Fig.~\ref{fig:mapping}. 
Hence, the evolution of the virtual state at DT is given by
\begin{equation} \label{equ:X_hat}
\widehat{X}_{n,t} (u_{n,t}) = \begin{cases} {X}_{n,t-D_{n,t}(u_{n,t})}, \ \ \ \ \ \ 
 \gamma_{n,t} =1,\\
\widehat{X}_{n,t-1}(u_{n,t-1}),    \ \ \ \ \ \
  \text{otherwise}. \end{cases}
\end{equation}
% One-dimensional state
{\color{black} In this context, DT synchronization directly requires reducing the mismatch between the physical and virtual states, rather than merely reducing transmission delay.
To evaluate DT synchronization, we calculate mismatch between the virtual state $\widehat{X}_{n,t}$ and the physical state $X_{n,t}$ of device $n$ at time slot $t$.}
% Given the virtual state $\widehat{X}_{n,t}$, we calculate the mismatch between $\widehat{X}_{n,t}$ and $X_{n,t}$ of device $n$ at time slot $t$.
% Given $\widehat{X}_{n,t}$, the mismatch between the virtual state and the physical state of device $n$ at time slot $t$ can be obtained. 
Since we consider two types of sensors: 1)  thermo-hygrometers  and 2) positioning sensors,
we use different loss functions to measure their mismatch. 
The loss function used to measure the temperature mismatch which extracts just noticeable differences~(JND)~\cite{Weber} for humans is 
\begin{equation} \label{equ:mismatch}
Z_{n,t}(u_{n,t}) =  \max \left\{
\frac{\left|X_{n,t}-\widehat{X}_{n,t}\right|}{\left|\widehat{X}_{n,t} \right| }
- \xi_n^{\mathrm{th}}, 0 \right \}, n \in \mathcal N_J, 
\end{equation}
% Z_{n,t}(u_{n,t}) = \max \left\{
% \frac{\left\|{X}_{n,t} - \widehat{X}_{n,t}(u_{n,t}) \right\|^2}{{X}_{n,t} ^2} - \xi_n^{\mathrm{th}}, 0 \right\} ,
where $\xi_n^{\mathrm{th}}$ is the threshold of device $n$ to filter the insignificant mismatch, $\mathcal N_J$ is the device set of thermo-hygrometers, and $u_{n,t}$ is omitted for notational simplicity.
The loss function used to measure positioning mismatch~\cite{positioning_mse} is 
\begin{equation} \label{equ:mismatch_}
Z_{n,t}(u_{n,t}) = \max \left\{
\left|X_{n,t}-\widehat{X}_{n,t}\right| -\xi_n^{\mathrm{th}} , 0 \right \} , n \in \mathcal N_P,
\end{equation}
where $\mathcal N_P$ is the device set of positioning sensors.
% Due to the complexity of the physical states, we invoke different types of mismatch function. 

\begin{comment}

In particular, for motion trajectory data we use
\begin{equation} \label{equ:mismatch}
Z_{n,t}(u_{n,t}) = \max \left\{
\frac{\left\|{X}_{n,t} - \widehat{X}_{n,t}(u_{n,t}) \right\|^2}{{X}_{n,t} ^2} - \xi_n^{\mathrm{th}}, 0 \right\} ,
\end{equation}
where $\xi_n^{\mathrm{th}}$ is the threshold of device $n$ to trigger the mismatch, and the denominator is used for normalization~\cite{Weber}. 

For high-dimensional state consists of media data such as the text and video, the mismatch can also be obtained by the cosine similarity between the extracted features of the physical state and virtual state~\cite{aoii_JSTSP}, given by 
\begin{equation}  \label{equ:mismatch_}
Z_{n,t}(u_{u,t}) =
\max \left\{
\frac{\mathbf{B}({X}_{n,t}) \cdot \mathbf{B}(\widehat{X}_{n,t}(u_{n,t}))^\top}
{\|\mathbf{B}({X}_{n,t})\| \cdot\|\mathbf{B}(\widehat{X}_{n,t}(u_{n,t}))\|} - \xi_n^{\mathrm{th}}, 0 \right\}, 
\end{equation}
where $\mathbf{B}(\cdot)$ is the pre-trained deep learning model to extract the features of media data.

% Multidimensional state
% \begin{equation} \label{equ:mismatch_}
% Z_{n,t} =\frac{\mathbf{B}({X}_{n,t}) \cdot \mathbf{B}({X}_{n,t})^\top}{\|\mathbf{B}(s)\| \cdot\|\mathbf{B}(\widehat{X}_{n,t})\|} ,
% \end{equation}
% where $\mathbf{B}$ is a discrimination function pre-trained encoder such as BERT, which maps multidimensional state to its vector space.
\end{comment}

From (\ref{equ:Bernoulli})-(\ref{equ:mismatch_}), we can see that the mismatch $Z_{n,t}$ depends on the physical state ${X}_{n,t}$, the device selection  decision $u_{n,t}$ and packet error rate $p_n$.
% the virtual device state  $\widehat{X}_{n,t}$ depends on the transmission decision $u_{n,t}$ and packet error rate $p_n$.
As $D_{n,t}$ and $p_n$ increase, the mismatch $Z_{n,t}$ increases. 
Therefore, to maintain the DT synchronization, the BS must guarantee the timeliness of sensing data transmission through the device scheduling policy adapting to physical state changes and dynamic transmission environments.

\color{black}
\subsection{Problem Formulation}

% \color{black}
% 考虑下资源分配对于 全局 AoII 的影响，资源分配是不是动态的？如何唤醒为终端的资源分配？
% \color{black}

{\color{black}To consistently maintain DT synchronization, we formulate an optimization problem whose goal is to minimize the weighted mismatch in DT system}, considering the diverse transmission priorities of heterogeneous devices (i.e., upheaval temperature data transmission is more urgent than positioning data). This problem is optimized by determining the device scheduling vector $\boldsymbol{u}_t$ 
% RB allocation policy ${\pi}$
at the BS, given by
% \begin{align}
% \mathbb{P} 1:   
% & \min_{ \pi} \ \overline{J}(\mathcal{N}, t, X_t) \label{equ:p_min_aoii} 
% % & \min_{ \pi}  \limsup_{T \rightarrow\infty} \frac{1}{TN} \mathbb{E}_{\tau \sim \pi} \left[ \sum_{t=0}^{T-1} \sum_{n=1}^{N} w_n \Delta^{\pi}\left( n,t \right) \mid \Delta(n,0)  \right] \label{equ:p_min_aoii} 
% \\    
% & \text { s.t. }  \
% u_{n,t} \in \{0,1\}, \quad t \in \mathcal{T}, \ \ n \in \mathcal{N},  \tag{\ref{equ:p_min_aoii}{a}}  \label{equ:p_min_aoii_a}
% \\
% % 0到T-1的写法
% % t= 0, \ldots, T-1
%  % \limsup _{T \rightarrow \infty} \frac{1}{T}  \sum_{t=0}^{T-1} 
% & \qquad  \quad \boldsymbol{b}_t^\top \boldsymbol{u}_t \leq M, \quad t \in \mathcal{T},
% \tag{\ref{equ:p_min_aoii}{b}} 
%    \label{equ:p_min_aoii_b}
% \end{align}
\begin{align}
\mathbb{P} 1:   
&  \min_{\boldsymbol{u}_t }  \limsup_{T \rightarrow \infty} \frac{1}{TN}  \mathbb{E}_{\tau }\left(
\sum ^{T}_{t=1} \boldsymbol{w}^{\top}   \boldsymbol Z_{t}(\boldsymbol{u}_t)   \right) 
\label{equ:p_min_aoii} 
\\    
& \text { s.t. }  \
u_{n,t} \in \{0,1\}, \quad    n \in \mathcal{N}, \ \ t \in \mathcal{T},  \tag{\ref{equ:p_min_aoii}{a}}  \label{equ:p_min_aoii_a}
\\
& \qquad  \lim_{T\rightarrow \infty }\frac{1}{T}   \sum ^{T}_{t=1} \mathbb{E}_{\tau }\left(  \boldsymbol b_{}^{\top}  \boldsymbol u_{t} \mid \boldsymbol Z_0 \right) \leq M, 
\tag{\ref{equ:p_min_aoii}{b}}  \label{equ:p_min_aoii_b}
% \\
% & \qquad \sum ^{N}_{n=1} b_n >M,
% \tag{\ref{equ:p_min_aoii}{c}}  \label{equ:p_min_aoii_c}
\end{align}
% \begin{align}
% \mathbb{P} 1: &  
% \min_{\pi \in \Pi}  \limsup _{T \rightarrow+\infty} \frac{1}{T} \mathbb{E}_{\tau \sim \pi}\left( \sum_{t=0}^{T-1} \Delta_{A o I I}\left(s_{n,t}^\pi \right) \mid s_{n,0}\right) \label{equ:prob_min_aoii}
% \\
% & \text { s.t. }  
% \limsup _{T \rightarrow+\infty} \frac{1}{T} \mathbb{E}_{\tau \sim \pi}\left( \sum_{t=0}^{T-1} \mu_{n,t}^\pi \mid s_{n,0}\right) \leq \delta_n,  \  \forall s_{n,0} \in \mathcal{I}_F,\tag{\ref{equ:prob_min_aoii}{a}} \label{equ:prob_min_aoii_a}
% \end{align}
where
$\boldsymbol u_{t} = \left[ u_{1,t}, \cdots, u_{N,t} \right] ^\top$ is the device scheduling vector;
$\boldsymbol{w} = \left[ w_1, \cdots, w_N \right] ^\top$ is the transmission priority vector of all devices with $w_n$ being the priority of the sensing data of device $n$;
$\boldsymbol{Z}_t(\boldsymbol{u}_t) = \left[Z_{1,t}(u_{1, t}), \cdots, Z_{N,t}(u_{N, t}) \right]^\top$ is the mismatch vector of all the devices at time slot $t$;
$\boldsymbol{Z}_0$ is the initial mismatch vector at time slot 0.
${\tau} = \left[ \boldsymbol{Z}^{}_1 (\boldsymbol u_1), \cdots, \boldsymbol{Z}^{}_{T}(\boldsymbol u_T) \right]$ is the evolution trajectory of the mismatch vector determined by $\boldsymbol u_t$;
% $\boldsymbol{\tau} = \left[ \tau_1, \cdots, \tau_N \right]$ is the trajectory of $\boldsymbol{Z}^{\pi}_t$; 
$\mathcal{T}$ is the set of all considered time slots;
$\mathcal{N} = \mathcal{N}_J \cup \mathcal{N}_P$ is the set of all devices;
% $\boldsymbol{u}_t=\left[u_{1,t}, \cdots, u_{N,t} \right]$ is the transmission decision vector of devices at time slot $t$, 
% $\tau = \left[\boldsymbol{s}_0, \cdots, \boldsymbol{s}_{T-1}^\pi\right]$ with $\boldsymbol{s}_t = \left[ s_{1,t}, \cdots, s_{N,t} \right]$. \color{red} $\pi$ is the policy to determine $\boldsymbol{u}_t$.
\begin{comment}   \color{black}
$w_n$ is the weight of device $n$ that indicates the significance of its sensing data, 
$\mathbb{E}_{\tau \sim \pi}(\cdot)$ is the expectation with respect to $\tau = \left[\tau_1, \cdots, \tau_N \right]$ generated by  $\pi$, \color{black}\end{comment}
$\boldsymbol{b}=\left[b_{1}, \cdots, b_{N} \right]^\top$ is the vector of the occupied RB numbers by each device in $\mathcal{N}$;
and $M$ is the total number of the RBs of the BS which is inadequate to schedule all devices at the same time. \color{black} Constraints (\ref{equ:p_min_aoii_a}) and 
(\ref{equ:p_min_aoii_b}) indicate that the BS can allocate at most $1$ RB to each device each time, and at most $M$ RBs to the scheduled devices overall.
% (\ref{equ:p_min_aoii_c}) indicates that the available RB number $M$ is inadequate to schedule all devices at the same time.
From (\ref{equ:p_min_aoii}), we can see that the device scheduling decision $\boldsymbol{u}_t$
% RB allocation policy $\pi$ 
and physical states $X_{t}$ determine the DT mismatch and the device scheduling decision $\boldsymbol{u}_t$ must satisfy the limited RB number constraint for the given $X_t$ and $\boldsymbol{b}$.

% indicates that all RBs occupied by all devices cannot exceed the limited RBs of the server at each time slot.

\begin{comment}

\color{red}
$\Delta^{\pi}\left( n,t \right)$ is the transitioned state at time slot $t$ under policy $\pi$, given an initial system state $s_{n,0}$.
$\mathbb{E}_{\tau \sim \pi} \left(\sum_{t=0}^{T-1}\Delta_{\mathrm{AoII}}\left(s_{n,t}^\pi\right) \mid s_{n,0}\right)$ is the expectation of long-term AoII for a trajectory of states $\tau_n = \left[s_{n,0}, \cdots, s_{n,T-1}^\pi\right]$ with $\mathbb{E}_{\tau \sim \pi}(\cdot)$ being the expectation of $\tau$ generated by policy $\pi$,
\color{black} 
\end{comment}

However, one can notice that the problem~(\ref{equ:p_min_aoii}) is challenging to solve via conventional optimization methods for the following reasons. 
First, from (\ref{equ:X_hat})-(\ref{equ:mismatch_}) we see that, since the evolution of mismatch $Z_{n,t}$ is determined by the sequence of 
device scheduling vector  $\boldsymbol{u}_t$, the problem~(\ref{equ:p_min_aoii}) belongs to 
{\color{black}CMDP} 
% constrained Markov decision process (CMDP)
\cite{CMDP_book} and is hard to be solved by conventional optimization methods~\cite{zyj_dt}. 
Second, the BS must know the mismatch vector $\boldsymbol Z_{t}$ of all devices to optimize device scheduling vector $\boldsymbol u_t$.
{\color{black}
However, $\boldsymbol Z_{t}$ is calculated at each device, hence the BS can only obverse partial $\boldsymbol Z_{t}$ from the scheduled devices, and cannot obtain $Z_{n,t}$ of the unscheduled devices.}
The partial observation makes the BS infeasible to solve the problem via optimization methods.
Third, the total available RB number $M$ may change with wireless environment dynamics during the prolonged DT mapping~\cite{enhanced_RL}.
However, conventional optimization methods do not consider variable $M$, and need to be re-implemented once $M$ changes. The dynamic programming~(DP) and standard reinforcement learning~(RL) methods can help learn device scheduling strategies in dynamic environments, but need to be enhanced in regards to their few shot performances with the ever-changing environments. 
Thus, we propose a CRL algorithm to learn and adjust device scheduling strategies 
% {\color{black} from regressed $\boldsymbol Z_t$},
with in-depth analysis on impacts of the constraints $M$ over the device scheduling vector $\boldsymbol u_t$, and quick adaptivity ability refined from historical experiences at multiple timescales.

\section{Proposed CRL Algorithm for CMDP}

% 拉格朗日转化是转化成无约束问题的(并且适合单步求解)，并利用 prime-dual 迭代原始-对偶方法 方法求解。
% 为了考虑安全性,（满足条件的性能）我们引入了state-wise的问题转化。
% 由于信源和M是随机变化的 unpredictable dynamic，我们采用了CRL的方法，来挖掘M对于policy 的影响。

In this section, we introduce the proposed CRL algorithm, which integrates MTR~\cite{muli_timescale_rep_CRL} into an SAC framework~\cite{fac}. 
%the proposed solution to solve the CMDP problem~(\ref{equ:p_min_aoii}).
% CMDP is usually processed via Lagrangian techniques, which transform the constrained problem to an unconstrained one and solve the transformed problem by prime-dual optimization~\cite{Dual_CMDP, CMDP_book, Primal_Dual_Methods_CMDP}. 
We will first transform the original long-term-constrained device scheduling problem~(\ref{equ:p_min_aoii}) into 
% one of finding prime-dual optimization form, thus considering a problem without constraint.
a Lagrangian dual problem, such that the goal of the considered system is to seek a prime-dual solution with refined constraints on optimization variables~\cite{CMDP_book}.
 % (state the purpose of transformation).
Then, the {\color{black} multiple timescale replay for soft actor critic (MTR-SAC)} enabled CRL algorithm will solve the non-convex dual problem considering the dynamic and unpredictable physical system status. 
An MTR buffer is integrated into the CRL algorithm to store experiences at different timescales, thus the CRL can learn 
% the impacts of changing constraints on device scheduling vector and learn 
a stable policy across the historical stages for adapting to unknown dynamics.
Meanwhile, the SAC framework enhances the action sampling efficiency with maximizing the logarithmic formatted action entropy.
% can select stochastic device scheduling vectors thus guaranteeing fairness. 
% The CRL algorithm introduced multi-timescale replay for soft actor-critic~(MTR-SAC) agent. 
% The CRL algorithm is deployed at the BS to centrally schedule devices for sensing data uploading. 
% {\color{red}{The CRL algorithm enables refining the impact of changing constraints on device scheduling vector, thus improving the performance of minimizing mismatches in dynamic network environments.}}
% device scheduling vector feature of 
% control the device scheduling vector according to constraint thus 
% to learn the impacts of the changing constraint on the policy, thus improving the performance of reducing mismatch with changing constraints
% The CRL algorithm aims to optimize the policy within the constraint through learning the impacts of the changing constraint on the policy.
% Hence, the proposed CRL algori't'h'm can improve the performance of reducing mismatch with changing constraints.
In what follows, we first explain how to transform the problem~(\ref{equ:p_min_aoii}) into a Lagrangian dual problem.
Then, we introduce components of the proposed CRL algorithm. 
Finally, we explain the training procedure of the proposed CRL algorithm.

% the Lagrangian dual problem is solved via a soft actor-critic~(SAC) based reinforcement learning~(RL) algorithm.
% Finally, we introduce multiple time-scale replay~(MTR) based continual learning scheme to the SAC agent to adapt the agent to changing RB number $M$. 

% \color{black}
% 这里可以考虑把原始问题的约束改成平均的，以做成 state-wise 的约束
% \color{black}
% analyse the define the components of the CMDP, then we perform the problem transformation, and finally we propose CRL based method to solve the transformed problem.

% \color{red}
% 求解的时候需要处理一下, 有没有分布式学习的必要性，如果终端是自主请求RB的话，终端之间没有必要调控相互的资源，利用集中式算法就能解决问题。基站利用集中请求，来获取终端的误差信息/适配的程度，从而集中调配。
% \color{black}
% To handle the constraint term, 
% we first transform the problem into a state-wise form via Lagrangian transform, which can be given by
\subsection{Lagrangian Transform}

Since constraint~(\ref{equ:p_min_aoii_b}) in problem~(\ref{equ:p_min_aoii}) is the expectation of the long-term consumed RBs, it may lead to temporary overbudget RB consumption, i.e. $\boldsymbol{b}_{t}^{\top} \boldsymbol u_{t}>M$. 
%the consumed RB number $\boldsymbol{b}_{t}^{\top} \boldsymbol u_{t}$ to temporary exceed $M$.
% temporary violation of constraint~(\ref{equ:p_min_aoii_b}) that the consumed RB number $\boldsymbol{b}_{t}^{\top} \boldsymbol u_{t}$ exceed $M$.
However, even a temporary resource overbudget can strain the available RBs and lead to congestion.
Such congestion in practice increases the data packet error rate thus breaking down the DT mapping.
% increasing the data packet error rate .
With such consideration,
we transform the expectation-wise constraint~(\ref{equ:p_min_aoii_b}) into a state-wise constraint~\cite{fac}, to maintain under-budget device scheduling.% vector does not exceed $M$ at each time slot.

\begin{comment}
which can cause the consumed RB number $\boldsymbol{b}_{t}^{\top}  \boldsymbol u_{t}$ to temporarily exceed $M$ at the early stage when solving problem~$\mathbb{P} 1$.
% the  problem~(\ref{equ:prob_min_aoii}). 
However, even a temporary exceedance of the constraint $M$ can strain the RBs and form congestion thus  increasing the data packet error rate.
To address this issue, 
we first transform the problem~(\ref{equ:p_min_aoii}) into a state-wise constrained problem to make the device scheduling vector satisfies constraint~(\ref{equ:p_min_aoii_b}) at each time slot.
% where the constraints can be easily handled, 
Then we obtain the Lagrangian dual problem of the state-wise form problem to solve~(\ref{equ:p_min_aoii}). 
\end{comment}

% \color{black}既然T是趋于无穷的，我们可以把每一个t的状态当做是初始状态。\color{black}
\subsubsection{State-wise Constrained Problem}
% The main idea of the state-wise form problem is to regard each state $\boldsymbol{Z}_t (\boldsymbol u_t)$ as an initial state in~(\ref{equ:p_min_aoii_b}), because the constraint~(\ref{equ:p_min_aoii_b}) considers $T\rightarrow \infty$.
% To this end, we formulate a state-wise constraint, which can guarantee each state satisfy the constraint, given by  

The state-wise constraint is given by~\cite{fac}
\begin{equation}
\begin{aligned}
    &V_c \left( \boldsymbol{Z}_t^{} (\boldsymbol{u}_t)\right) = \\
    &\lim_{T \rightarrow \infty} \frac{1}{T} \mathbb{E}_{\tau }\left(\sum_{t=1}^{T} \boldsymbol{b}^{\top} \boldsymbol{u}_t^{} \mid \boldsymbol{Z}_0=\boldsymbol {Z}_t^{}(\boldsymbol{u}_t)\right) \leq M, \forall  \boldsymbol {Z}_t^{} \in \mathcal{I}_F, \label{equ:def_V_c}
\end{aligned}
\end{equation}
where  $\mathcal{I}_F = \mathcal{I} \cap \mathcal{S}_F$ is the feasible region, with $\mathcal{I}$ being the set of all possible initial mismatch vector $\boldsymbol{Z}_0$ and $\mathcal{S}_F$ being the feasible set of $\boldsymbol {Z}_t^{}$.
Compared to constraint~(\ref{equ:p_min_aoii_b}), the state-wise constraint~(\ref{equ:def_V_c}) regards each mismatch vector $\boldsymbol{Z}_t $ as an initial point in~(\ref{equ:p_min_aoii_b}), because the constraint~(\ref{equ:p_min_aoii_b}) considers $T\rightarrow \infty$.
Hence, the problem with state-wise constraint is 
% Hence, without changing the objective, 
% we reformulate the original problem $\mathbb{P} 1$ with state-wise constraint, which is given by
% we leverage the property of Lagrange multiplier $\lambda$ of $\mathbb{P} 1$, which indicates the importance of constraints in.
% In particular, we assign a Lagrange multiplier to each $s_t$, to ensure that the constraint is always satisfied.
% % \color{black}
% Consequently, the problem $\mathbb{P} 2$ can be reformulated into a state-wise constrained problem, which is given by
% \begin{align}
% \mathbb{P} 2:  \ & \min _{\boldsymbol\tau \in \Pi}  \limsup_{T \rightarrow \infty} \frac{1}{TN} \mathbb{E}_{\boldsymbol\tau  \sim \boldsymbol\pi}\left(\sum_{t=0}^{T-1} \boldsymbol{w}^{\top} \boldsymbol{Z}_t^\boldsymbol\pi\right) \label{equ:prob_2}
% \\
% & \text { s.t. } V_c\left(\boldsymbol{b}, \boldsymbol{Z}_t^\boldsymbol\pi\right) \leqslant M, \ \forall \boldsymbol{Z}_t^{\boldsymbol\pi}  \in \mathcal{I}_F,\tag{\ref{equ:prob_2}{a}}  \label{equ:prob_2a} 
% \\
% & \qquad V_c\left(\boldsymbol{b}, \boldsymbol{Z}_t^\boldsymbol\pi\right)
% =\lim _{T \rightarrow \infty} \frac{1}{T} \mathbb{E}_{\boldsymbol\tau  \sim \boldsymbol\pi}\left(\sum_{i=0}^{T-1} \boldsymbol{b}^{\top} \boldsymbol{u}_t^\pi \mid Z_0=Z_t^\pi\right),
% \tag{\ref{equ:prob_2}{b}}  \label{equ:prob_2b}
% \end{align}\label{equ:prob_2b_} 
\begin{align}
\mathbb{P}2: \quad & \min _{\boldsymbol u_t} \limsup_{T \rightarrow \infty} \frac{1}{TN} \mathbb{E}_{\tau }\left(\sum_{t=1}^{T} \boldsymbol{w}^{\top} \boldsymbol{Z}_t^{}(\boldsymbol u_t)\right) \label{equ:prob_2} 
\\
& \text{s.t. } \ V_c \left(\boldsymbol{Z}_t^{}(\boldsymbol u_t)\right) \leq M, \ \forall \boldsymbol{Z}_t^{} \in \mathcal{I}_F.
\tag{\ref{equ:prob_2}{a}} \label{equ:prob_2a} 
% \\
% & \qquad V_c\left(\boldsymbol{b}, \boldsymbol{Z}_t^{\boldsymbol\pi}\right) = \lim_{T \rightarrow \infty} \frac{1}{T} \mathbb{E}_{\boldsymbol\tau \sim \boldsymbol\pi}\left(\sum_{i=0}^{T-1} \boldsymbol{b}^{\top} \boldsymbol{u}_t^{\pi} \mid Z_0=Z_t^{\pi}\right), 
\end{align}
% where $V_c \left(\boldsymbol{b}, \boldsymbol{Z}_t^{\boldsymbol\pi}\right)$ is a function that evaluates the temporary consumed RBs, given by 
% where the constraint~(\ref{equ:p_min_aoii_c}) is omitted due to the irrelevance with $\boldsymbol u_t$.
Since the state-wise constraint~(\ref{equ:prob_2a}) takes each mismatch vector $\boldsymbol {Z}_t(\boldsymbol u_t)$ as an initial point, the consumed RB number $\boldsymbol{b}_{t}^{\top} \boldsymbol u_{t}$ temporarily exceeding $M$ violates the constraint~(\ref{equ:prob_2a}).
{
\color{black} In this way, $\mathbb{P} 2$ has a stricter constraint than $\mathbb{P}1$ to guarantee under budget resource consumption, and the optimal solution of $\mathbb{P} 2$ will be sub-optimal to  $\mathbb{P}1$.
However, considering the fact that the optimal solution of $\mathbb{P}1$ can lead to a temporary overconsumption of resources, the optimal solution of
$\mathbb{P} 2$ will be the one that identifies the practically feasible optimal solution.
}

% From (\ref{equ:def_V_c}) we see that, $V_c\left(\boldsymbol{Z}_t^{}(\boldsymbol u_t)\right)$ regards each state $\boldsymbol {Z}_t^{}$ at time slot $t$ as a initial state $\boldsymbol{Z}_0$, hence (\ref{equ:prob_2a}) becomes a state-wise constraint. 
% Thus, the impact of the early stage $\boldsymbol {Z}_t^{}$ is accounted to the RB constraint. 
% In this way, $\mathbb{P}2$ has a stricter constraint than $\mathbb{P}1$ to guarantee the temporary constraint and the solution of $\mathbb{P}2$ is a sub-optimal solution to  $\mathbb{P}1$.

\subsubsection{Lagrangian Dual of State-wise Constrained Problem}

% We use the Lagrangian method to process problem $\mathbb{P}2$. 
Since the state-wise constraint~(\ref{equ:prob_2a}) forces a constraint for each $\boldsymbol {Z}_t$ in $\mathcal{I}_F$, we employ one corresponding Lagrangian multiplier on each $\boldsymbol {Z}_t$ and denote $\lambda (\boldsymbol {Z}_t )$ as the multiplier of each $\boldsymbol {Z}_t$~\cite{fac}.
Therefore, taking into account $\boldsymbol{Z}_t^{} \in \mathcal{I}_F$, the 
% Lagrangian function of problem $\mathbb{P}2$, named by 
original state-wise Lagrangian function $\mathcal{L}_{\text{o-stw}}(\boldsymbol u, \lambda)$ of problem $\mathbb{P}2$ 
is 
\begin{equation} \label{equ:o-stw}
\begin{aligned}
\mathcal{L}_{\text{o-stw}}(\boldsymbol u, \lambda) =\limsup _{T \rightarrow \infty} & \frac{1}{T} 
\mathbb{E}_{\tau}
\left(\sum \frac{1}{N} \boldsymbol{w}^{\top} \boldsymbol{Z}_t^{}(\boldsymbol u_t)\right)+ \\
&\sum_{\boldsymbol{Z_t} \in \mathcal{I}_F} \lambda\left(\boldsymbol{Z}_t^{{}} (\boldsymbol u_t)\right) \Big[V_c\left(\boldsymbol{Z}_t (\boldsymbol u_t)\right)-M\Big],
\end{aligned}
\end{equation}
where $\lambda\left(\boldsymbol{Z}_t^{} (\boldsymbol u_t) \right)$ is the state-wise multiplier for each $\boldsymbol{Z}_t^{{}}$.
% and $\boldsymbol{Z}$ is the feasible mismatch vector of all devices.
Different from the traditional Lagrangian method that assigns one $\lambda$ to all $\boldsymbol {Z}_t$~\cite{Aoii_TWC}, $\mathcal{L}_{\text{o-stw}}(\boldsymbol u, \lambda)$ applies different Lagrangian multipliers to each $\boldsymbol {Z}_t$, with which the infeasible $\boldsymbol {Z}_t$ values can be indicated.
%Since the Lagrangian multiplier indicates which  $\boldsymbol {Z}_t$ is infeasible, the setting of  $\lambda\left(\boldsymbol{Z}_t^{{}} (\boldsymbol u_t)\right)$ enhance the constraints on $\boldsymbol u_t$.

Notice that it is hard to demarcate the feasible region $\mathcal{I}_F$, while impractical to find the summation of infinite set $\mathcal{I}_F$, 
%In (\ref{equ:o-stw}), it is hard to demarcate the feasible region $\mathcal{I}_F$. Besides, calculating the summation of infinite set $\mathcal{I}_F$ is impractical. 
% This leads to the existing solutions for CMDP rarely considering the state-wise constraint.
In such context, we use experience sampling based learning method to transform  $\mathcal{L}_{\text{o-stw}}(\boldsymbol u, \lambda)$ into an equivalent form $\mathcal{L}_\text{stw}(\boldsymbol u, \lambda)$, which is given by 
\begin{equation}
\begin{aligned} \label{equ:stw}
&\mathcal{L}_\text{stw}(\boldsymbol u, \lambda)= \\
% &\sum_{\boldsymbol{Z} \sim d_0(\boldsymbol{Z})}\left[\limsup _{T \rightarrow \infty}  \frac{1}{TN} \sum_{t=0}^{T-1} \boldsymbol{w}^{\top} \boldsymbol{Z}_t^\pi+\lambda\left(\boldsymbol{Z}_t^{{\pi}}\right) \left(V_c\left(\boldsymbol{Z}_t^{{\pi}}\right)-M \right) \right], \\
&\mathbb{E}_{\boldsymbol{Z} \sim d_0(\boldsymbol{Z})} \left[  \frac{1}{N} \boldsymbol{w}^{\top} \boldsymbol{Z}_t(\boldsymbol u_t)+\lambda\left(\boldsymbol{Z}_t(\boldsymbol u_t)\right) \left(V_c\left(\boldsymbol{Z}_t(\boldsymbol u_t)\right)-M \right)  \right],
\end{aligned}
\end{equation}
where $d_0(\boldsymbol{Z})$ is the statistical distribution of $\boldsymbol Z_t$ obtained by  sampling method.
The proof of the equivalence of optimizing $\mathcal{L}_{\mathrm{stw}}( \boldsymbol u, \lambda)$ and $\mathcal{L}_{\text {o-stw }}( \boldsymbol u, \lambda)$ is illustrated by \textit{\textbf{Theorem 1}}.
\\

\textbf{\textit{Theorem 1 }}
% (Equivalence of $\mathcal{L}_{\mathrm{stw}} (\boldsymbol\pi, \lambda)$ and $\mathcal{L}_{\text {o-stw }}(\boldsymbol \pi, \lambda)$ ):
If the  Lagrangian dual problem of $\mathcal{L}_{\mathrm{stw}}( \boldsymbol u, \lambda)$, which is $\max_{\lambda} \inf_{\boldsymbol u} \mathcal{L}_{\text {stw}}(\boldsymbol u, \lambda)$, has the optimal scheduling vector sequence $\boldsymbol u^*$ and Lagrange multiplier $\lambda^*$,  then $\boldsymbol u^*$ is also the optimal scheduling vector sequence for the state-wise constrained problem $\mathbb{P} 2$. 

% If the pair of the optimal policy $\pi^*$ and Lagrange multiplier $\lambda^*$ exist for  problem $\max_{\lambda} \inf_{\pi} \mathcal{L}_{\text {stw}}(\pi, \lambda)$, then $\pi^*$ is also the optimal policy for the state-wise constrained problem $\mathbb{P} 2$. 

\textit{Proof:} See Appendix A.   $\hfill\blacksquare$
\\
% The idea is to scale the constraints by the probabilistic density $d_0(\boldsymbol{Z}_t)\left(\boldsymbol{b}^\top \boldsymbol{u}_t |\boldsymbol{Z}_t- M \right) \leq 0$. Since $d_0(\boldsymbol{Z}_t) \geq 0$, and for an infeasible initial state $\boldsymbol{Z}_t \notin \mathcal{I}_F$, $d_0(\boldsymbol{Z}_t)=0$, so the scale constraint is equivalent to the state-wise constraint in $\mathbb{P} 3$. 

% For the infeasible initial states ${s} \notin \mathcal{I}_F$, $\lambda$ goes to infinity.     

With \textit{\textbf{Theorem 1}}, the equivalent dual problem of $\mathbb{P} 2$ is given by~\cite{Primal_Dual_Methods_CMDP}
% the prove of equality between $\mathcal{L}_\text{stw}(\boldsymbol\pi, \lambda)$ and $\mathcal{L}_{\text{o-stw}}(\boldsymbol\pi, \lambda)$ is given by \color{red} XX \color{black}.
% The dual problem is given by  
\begin{align}
 \mathbb{P} 3: & \max _\lambda \inf_{ \boldsymbol u} \mathcal{L}_{\text{stw}}( \boldsymbol u, \lambda)  \qquad \text{s.t.} \lambda \geq 0.
\end{align}
% \color{black}
% Therefore, the solution of $\mathbb{P}_3$ is feasible for $\mathbb{P}_2$.
% From the prove we see that, $\mathbb{P}_3$ is equivalent to $\mathbb{P}_2$. \color{black}
% 因为是信源是随机变化的，而且约束也在变，所以用动态规划的技术不好解决，因此，我们使用CRL来学习。
Since we consider the real-word physical states $X_t$ change with unpredictable dynamics, 
$\boldsymbol{Z}_t(\boldsymbol u_t)$ is usually non-convex to $\boldsymbol u_t$, 
making it hard to solve the problem $\mathbb{P} 3$ by convex optimization or 
% model the transition probability of mismatch $\boldsymbol Z_{t}$ using 
DP approach.
Moreover, the changing of $M$ leads to varying environments that severely affects the performance of standard RL methods.
Hence, to solve problem $\mathbb{P} 3$, we employ a CRL algorithm to learn a device scheduling policy $\boldsymbol u$ and the multiplier function  $\lambda (\boldsymbol{Z}_{t})$  which adapts to dynamic $X_t$ and $M$ via learning from the sampled experiences.
% and $\lambda(\boldsymbol{Z}_{t}(\boldsymbol u_t))$.

\vspace{-0.3cm}
\subsection{Components of the CRL Algorithm}
% The CRL algorithm uses multi-timescale replay~(MTR) scheme for soft actor-critic~(SAC) agent. 
% The CRL algorithm uses multi-timescale replay for soft actor-critic~(MTR-SAC) agent.
The proposed CRL algorithm uses multi-timescale replay for soft actor-critic~(MTR-SAC) agent.
% by introducing multi-timescale replay for soft actor-critic~(SAC) agent.
A standard SAC model consists of two important modules,  actor-critic network module, and replay buffer module. 
With these two modules, SAC can be trained with temporal-difference~(TD) error and experience replay, respectively, thus achieving high learning efficiency with limited training data~\cite{Soft_Actor_Critic}. 
Moreover, through maximizing action entropy, SAC learns a more stochastic policy compared to other standard RL such as deep Q-network~(DQN)~\cite{Soft_Actor_Critic}.
MTR is introduced as a data rehearsal scheme that balances the amounts of long-term and short-term memory in the replay buffer, thus realizing continual learning across dynamic environments~\cite{muli_timescale_rep_CRL}.
Next, we will introduce components of the proposed MTR-SAC method.
% Then, we will explain its training process.

% To solve the problem~$\mathbb{P} 3$ with RL adapting to the changes of $\boldsymbol{Z}_{t}$, we define the components of the CMDP, including:
The MTR-SAC method consists of the following components:
1) Agent, 2) State, 3) Action, 4) Policy, 
5) Reward, 6) Cost, 7) MTR buffer, and 8) Multiplier function, which are elaborated as follows:
\begin{itemize}
    \item \textbf{Agent:} The agent performs the proposed method is the BS that consecutively schedule devices for uplink sensing data transmission. 
    
    %select the subset of devices and allocate the required RBs to the scheduled devices to support sensing data uplink transmission.
    
\begin{figure}[t]
    \centering
    \includegraphics[width=1.\linewidth,
    % height = 0.22\textwidth%,bb= 50 50 520 480
    ]{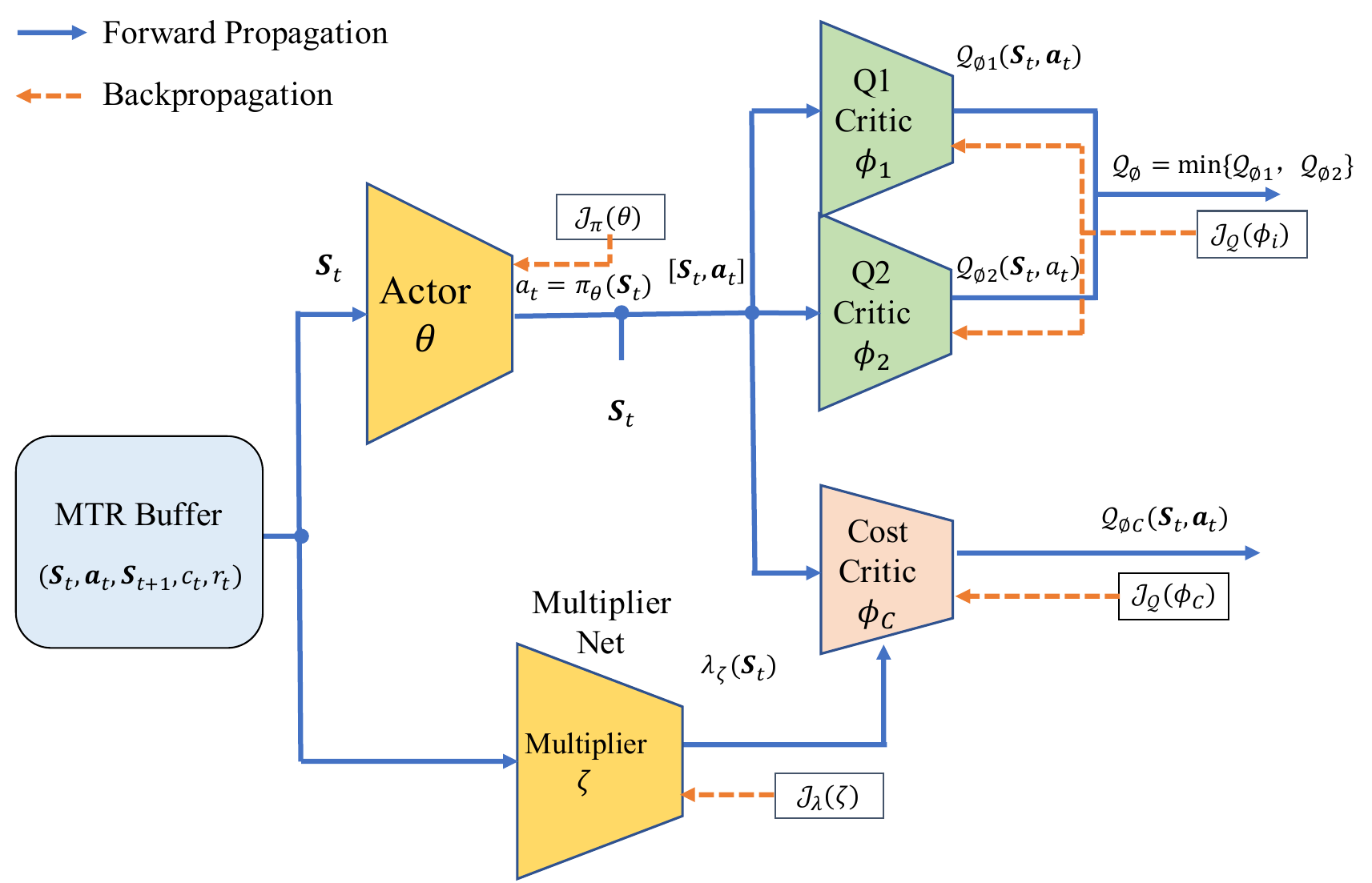}
    \setlength{\abovecaptionskip}{-0.cm}
    \caption{Architecture of the proposed MTR-SAC method. }\label{fig:SAC}
    \vspace{-0.cm}
\end{figure}

    \item \textbf{State:} The state at the BS is defined as $\boldsymbol{S}_t=\left[ \boldsymbol{s}_{1,t} , \cdots, \boldsymbol{s}_{N,t}\right]$, where $\boldsymbol{s}_{n, t} = \left[ \varPhi_{n,t}\left(u_{n, t}\right), 
    Y_{n,t},  \gamma_t \right]$
     is the status of each device $n$ at time slot $t$,  with  $\varPhi_{n,t}\left(u_{n, t}\right) = t- g_{n,t}^{\mathrm{m}}$ being the time span since the last time when data from device $n$ is correctly received by the BS, $g_{n,t}^{\mathrm{m}} = \sup \{t \mid \gamma_{n,t} =1 \}$ being the latest time slot when the data from device $n$ is correctly accepted, and $Y_{n,t} = Z_{n,g_{n,t}^{\mathrm{m}}}$ being the latest received mismatch value.
    % , which is determined by $u_{n, t}$.
    % AoI at BS of the sensing data transmitted by device $n$.
    Notice that,  $\varPhi_{n,t}\left(u_{n, t}\right)$ and $\gamma_{n, t}$ can be calculated at the BS, while $Y_{n,t}$ will be transmitted to the BS along with the sensing data.
    %In this way, the whole defined state of all devices $\boldsymbol{s}_t$ can be observed by the BS.

    % The state of the CMDP combines the AoI at the server, estimation error, and the data reception indicator, which is given by $\boldsymbol S_t = \left[ \Delta_{\mathrm{age}}(t), g(X_t,\hat{X}_t), \gamma_t \right]$.
    % is $\boldsymbol S_t\in \mathbb{N}$, which indicate AoII at time $t$
    \item \textbf{Action:} The action of BS is the scheduling index $u_{n,t}$ for each device $n$, at each time slot $t$. In particular, $u_{n,t}=1$ indicates the device $n$ is scheduled to  transmit data with $b_n$ allocated RBs at time slot $t$, while $u_{n,t}=0$ indicates that the device $n$ remains idle at time slot $t$. The vector of BS's actions at time slot $t$ is captured in $\boldsymbol a_t=[u_{1,t}, \cdots, u_{N,t}]^\top$.

% The agent takes an action $\boldsymbol a_t
%     % = \boldsymbol u_t
%     $ 
%     based on the current state $\boldsymbol S_t$, where $\boldsymbol a_t=[u_{1,t}, \cdots, u_{N,t}]^\top$, and $u_{n,t}$ is the scheduling decision for device $n$. 
%     $u_{n,t}$ determines whether the device $n$ is scheduled to transmit the sensing data or not at time slot $t$.
%     % samples and transmits data at time slot $t$ .
%     % $\pi=\left[\mu_0, \ldots\mu_t\right]$ to decide whether to sample and send the data at time $t$. 
%     $u_{n,t}=1$ indicates the device $n$  transmits data with the allocated $b_n$ RBs at time slot $t$, while $u_{n,t}=0$ indicates that the device $n$ remains idle at time slot $t$.
 
    % \color{orange}  In RL algorithm, $c(\boldsymbol{S}_t, \boldsymbol a_t)$ is the probability that an $\boldsymbol a_t$ violates the constraint.\color{black}

    \item \textbf{Policy:} The policy $  \pi_\theta \left( \boldsymbol a_t  \mid \boldsymbol S_t\right)$ is the probability that the BS chooses an action $\boldsymbol a_t$ at state $\boldsymbol{S}_t$, where 
    the policy is implemented at the actor network with a neural network (NN)  parameterized by $\theta$, as shown in Fig.~\ref{fig:SAC}.% that learns the relationship among the physical states, device scheduling vector and the DT mismatch.

% The policy is implemented by the DNN with parameter ϕv, which establishes the relation between the semantic triple scores, the ISS, and the transmission latency of each user. 
% Then, the conditional probability of each agent taking action αv in a given partial state sv can
% be expressed as πϕv (αv | sv). 

% To improve the action
% exploration, the policy networks are trained to maximize
% not only the expected reward, but also the entropy of
% actions H(πϕv (αv | sv)) which drives the agent to
% choose actions more randomly.

\begin{figure}[t]
    \centering
    \includegraphics[width=0.75\linewidth,
    % height = 0.22\textwidth%,bb= 50 50 520 480
    ]{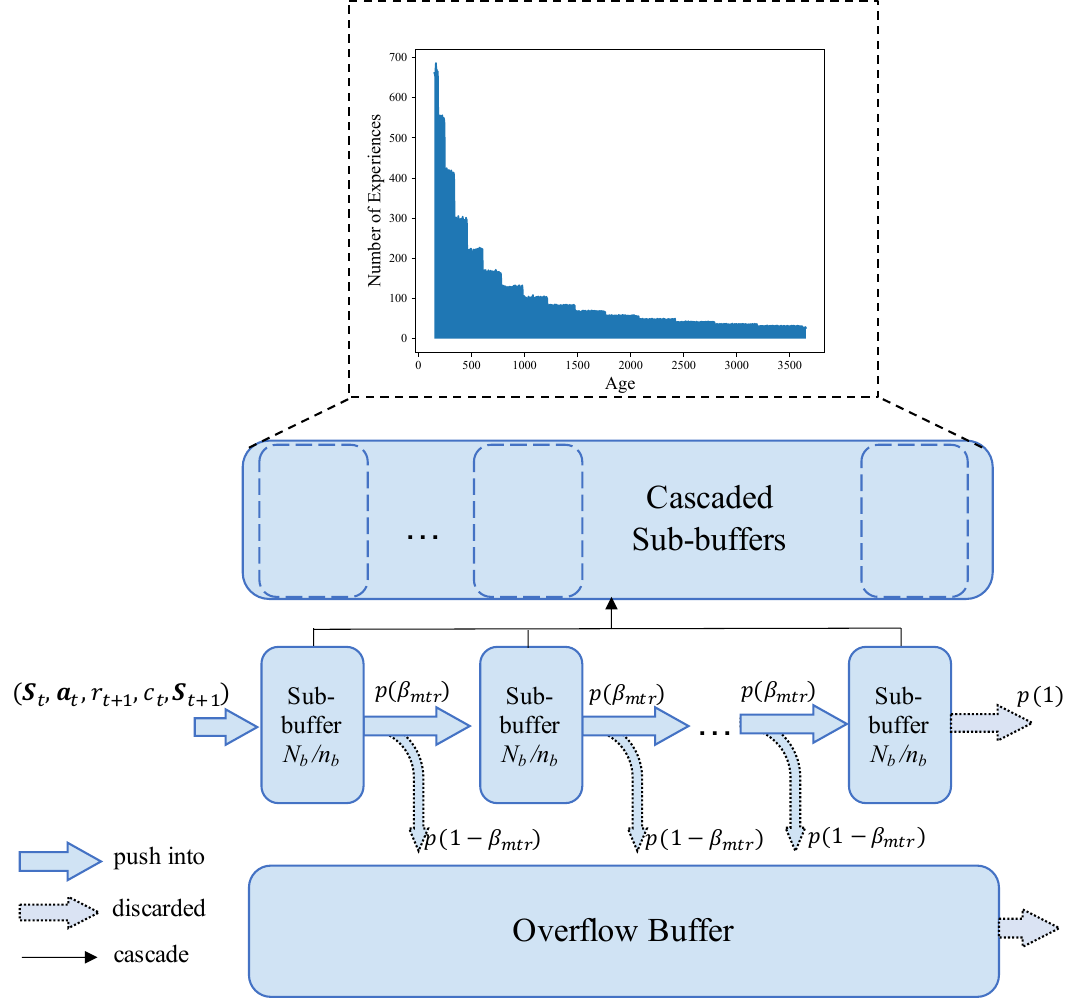}
    \setlength{\abovecaptionskip}{0.cm}
    \caption{\color{black}Architecture of MTR buffer. }\label{fig:MTR}
\vspace{-0.2cm}
\end{figure}

    \item \textbf{Reward:} The reward is used to capture the mismatch of all devices in the DT system  when a device scheduling action $\boldsymbol a_t$ is operated at state $\boldsymbol{S}_t$. Since the BS schedules devices and allocate RBs centrally, we define a reward function of the whole DT system as 
    \begin{equation}\label{equ:reward_def}
    \begin{aligned}
    r(\boldsymbol{S}_t, \boldsymbol{a}_t) &= -  \frac{1}{N} \boldsymbol{w}^{\top} \boldsymbol{Z}_t(\boldsymbol u_t) \\
    &= - \frac{1}{N} \sum_{n=1}^{N} w_{n} Z_{n,t} ( u_{n,t}),
    \end{aligned}
    \end{equation}
    where $w_n$ is the transmission priority weight of device $n$. 
    From (\ref{equ:reward_def}) we see that the reward $r(\boldsymbol{S}_t, \boldsymbol{a}_t)$ increases as the mismatch decrease, hence maximizing the reward leads to minimized weighted mismatch in the DT.
    {\color{black}Since the BS can only observe partial $\boldsymbol{Z}_t$ from the scheduled devices, the complete $\boldsymbol{Z}_t$ is supplemented with a certain delay after subsequent data collection by BS and is stored in the replay buffer specifically for the model training. }
    As shown in Fig.~\ref{fig:SAC}, the reward in the proposed method is regressed by  NN based reward value functions $Q_{\phi_i}\left(\boldsymbol S_t, \boldsymbol a_t\right), i \in \{1,2\}$ at the critic networks, where  $\phi_i$ is the NN parameters.

    \item \textbf{Cost:} 
    The cost captures the consumed RB number when a scheduling action $\boldsymbol a_t$ is operated at state $\boldsymbol{S}_t$. Consider the RB number constraint $M$, the cost is defined as 
    % The cost captures
    % the highest violation level on the RB number constraint, when a scheduling action $\boldsymbol a_t$ is operated at state $\boldsymbol{S}_t$, and is defined as
    % when a scheduling action $\boldsymbol a_t$ is operated at state $\boldsymbol{S}_t$, and is defined as   
    % The cost of the agent indicates the level of dissatisfaction regarding the constraints
    % % reflects the constraints of the CMDP 
    % and is given by 
    \begin{equation}
    c(\boldsymbol{S}_t, \boldsymbol{a}_t) = \begin{cases} M, \ \ \ \ \ \ 
     \boldsymbol{b}^\top \boldsymbol{a}_t \leq M,\\
    \boldsymbol{b}^\top \boldsymbol{a}_t,    \ \ \ 
    \boldsymbol{b}^\top \boldsymbol{a}_t > M .
    \end{cases}
          % c(\boldsymbol{S}_t, \boldsymbol{a}_t) = \max \left\{ \boldsymbol{b}^\top \boldsymbol{a}_t, ~M \right\}.
    \end{equation} 
    Based on this definition, the extent of an action 
    violating the RB constraints is represented by $c(\boldsymbol{S}_t, \boldsymbol{a}_t)-M$.
    The purpose of the cost definition is to evaluate the penalty of violating constraints separately from reward, so that the changing constraints do not influence the reward formula.
    The cost in the proposed method is regressed by NN based cost value functions  $Q_{\phi_C}\left(\boldsymbol S_t, \boldsymbol a_t\right)$ at cost critic network, where $\phi_C$ is the NN parameters, as shown in Fig.~\ref{fig:SAC}.

    \item \textbf{MTR buffer:} 
    The MTR buffer stores the experiences at different timescales and consists of $n_b$ cascaded first in first out~(FIFO) sub-buffers and a separate overflow buffer, as shown in Fig.~\ref{fig:MTR}.
    % which has a dynamic maximum size that is equal to the difference between $N$ and the number of experiences currently stored in the cascade (Figure 1(a)). 
    Any new experiences $\left(\boldsymbol S_t, \boldsymbol a_t, r_{t}, c_{t}, \boldsymbol S_{t+1}\right)$ are first pushed into the first sub-buffer. 
    As long as a sub-buffer is full, the oldest experience in the sub-buffer is popped out, then be pushed into the next sub-buffer with probability $\beta_{mtr}$ or be discarded to the overflow buffer with probability $1-\beta_{mtr}$.
    The overflow buffer is used to stabilize the total number of experiences in MTR buffer when the cascaded sub-buffers are not full in the early training stage.
    When the overflow buffer has $N_b$ discarded experiences, the overflow buffer is full and shrunk with the oldest experiences being thoroughly removed, until the whole MTR buffer has at most $N_b$ experiences.
    After all cascaded sub-buffers are full, the overflow buffer will be empty, and the distribution of the experience age in the cascaded sub-buffers results in a power law distribution of memories~\cite{muli_timescale_rep_CRL}.
    Hence, the samples of the experiences in MTR buffer can balance multi-timescale environment characteristic with containing abundant fresh experiences and a few old experiences. 
     % Once the cascade of sub-buffers is full, the size of the overflow buffer will be zero and any experience that is pushed out of any of the sub-buffers is discarded. 
     % During training, the number of experiences sampled from each sub-buffer (including the overflow buffer) is proportional to the fraction of the total number of experiences in the database contained in the sub-buffer. 
     % Figure 1(b) shows the distribution of ages of experiences in the MTR buffer, and a mathematical intuition for how the MTR buffer results in a power law distribution of memories is given in Appendix A.3.

    \item \textbf{Multiplier function:} 
    The multiplier function is the mapping relationship from each state $\boldsymbol S_t$ to the corresponding Lagrange multiplier value~\cite{fac}. 
    The multiplier function is implemented at the multiplier network with a NN and can be defined as $\lambda_{\zeta}\left(\boldsymbol S_t\right)$,  where $\zeta$ is the NN parameters.
    Given the definition, the BS can use experience samplings to obtain a estimation of $\mathcal{L}_\text{stw}(\boldsymbol u, \lambda)$ in~(\ref{equ:stw}), given by
    \begin{equation}
    \widehat{\mathcal{L}}_\text{stw} = \mathbb{E}_{\boldsymbol{S}} 
    \left[  
    r(\boldsymbol{S}_t, \boldsymbol{a}_t) -
    \lambda_{\zeta}\left(\boldsymbol S_t\right)
    \left(c\left(\boldsymbol{S}_t,\boldsymbol a_t)\right)-M \right)  \right],
    \end{equation}
    where $\mathbb{E}_{\boldsymbol{S}} (\cdot)$ is the expectation taken over the sampled $\boldsymbol S_t$ in MTR buffer. Then the MTR-SAC method is trained to maximize $\widehat{\mathcal{L}}_\text{stw}$.

    \begin{comment}
    \item \textbf{SAC module:} The SAC module uses neural networks to learn the Lagrange multiplier function $\lambda_{\zeta}\left(\boldsymbol S_t\right)$,
    % the stochastic policy and  comprises 
    two reward value functions $Q_{\phi_i}\left(\boldsymbol S_t, \boldsymbol a_t\right), i \in \{1,2\}$, one cost value function $Q_{\phi_C}\left(\boldsymbol S_t, \boldsymbol a_t\right)$, and the policy function $\pi_\theta\left(\boldsymbol S_t\right)$, where   $\zeta$, $\phi_i$, $\phi_C$, and $\theta$ is the corresponding parameters of these functions, respectively.
    The architecture of the proposed SAC module is illustrated  in Fig.~\ref{fig:SAC}. 
    In contrast to the standard SAC \cite{Soft_Actor_Critic},  a new neural network function $\lambda_{\zeta}\left(\boldsymbol S_t\right)$ is introduced to evaluate the multiplier $\lambda$ of $\boldsymbol S_t$ according to~(\ref{equ:o-stw}). 
    In this way, using experience samplings, the policy function $\pi_\theta\left(\boldsymbol S_t\right)$ can learn from an estimation of  $\mathcal{L}_\text{stw}(\boldsymbol u, \lambda)$ in~(\ref{equ:stw}), given by 
    % the $\mathcal{L}_\text{stw}(\boldsymbol u, \lambda)$ in~(\ref{equ:stw}) can be estimated by
    \begin{equation}
    \widehat{\mathcal{L}}_\text{stw} = \mathbb{E}_{\boldsymbol{S}} 
    \left[  r(\boldsymbol{S}_t, \boldsymbol{a}_t) 
    -
    \lambda_{\zeta}\left(\boldsymbol S_t\right)
    \left(c\left(\boldsymbol{S}_t,\boldsymbol a_t)\right)-M \right)  \right].
    \end{equation}
    \end{comment}

\end{itemize}
\vspace{-0.2cm}

\subsection{Training of the MTR-SAC Method }

Given the above MTR-SAC components, we next introduce the entire procedure of training the proposed MTR-SAC method. We first define the loss functions of the critic,  actor, and multiplier networks within the proposed MTR-SAC framework. Then we explain how this CRL solution is trained.

\subsubsection{Loss function of value (critic) networks}
% The original SAC actor network loss function without IRM scheme is  $J^{SAC}_\pi(\theta)$ as shown in~(\ref{equ:J_theta}), where $\alpha$ is the weight of action entropy, $Q_\phi\left(\boldsymbol S_t, \boldsymbol a_t\right)=\min \{Q_{\phi_1}\left(\boldsymbol S_t, \boldsymbol a_t\right), Q_{\phi_2}\left(\boldsymbol S_t, \boldsymbol a_t\right)\}$ is the Q value derived from the two reward value functions, 
% $\lambda_{\zeta}\left(\boldsymbol S_t\right)$ is the estimated multiplier value, and
% $ Q_{\phi_C}\left(\boldsymbol S_t, \boldsymbol a_t \right)$ is the estimated cost value.
The loss function to train the two reward value functions $Q_{\phi_i}\left(\boldsymbol S_t, \boldsymbol a_t\right)$ is given by~\cite{Soft_Actor_Critic}
\begin{align}
    J_Q\left(\phi_i \right)&=\mathbb{E}_{\left( \boldsymbol S_t, \boldsymbol a_t\right) \sim \mathcal{B}} \left[ \frac{1}{2} \Big( Q_{\phi_i}\left(\boldsymbol S_t, \boldsymbol a_t\right) \right.  \nonumber \\
    &  \left. - \mathbb{E}_{ \boldsymbol  a_{t+1} \sim \pi} \left(r\left(\boldsymbol S_t, \boldsymbol a_t\right)+ \gamma Q_{\overline{\phi}_i}\left(\boldsymbol S_{t+1}, \boldsymbol  a_{t+1}\right) \right) \Big)^2\right], 
    \label{equ:J_Q}
\end{align}
where $ i \in \{1,2\}$ is the index of the critic networks as show in Fig.~\ref{fig:SAC};  $\mathcal{B}$ is the MTR replay buffer; $\overline{(\cdot)}$ is the target network with a lower update frequency than the original network~\cite{SAC_algs_apps}; and $\gamma$ is the reward discounted factor. 
The stochastic gradient of $J_Q\left(\phi_i \right)$ can be given by
\begin{align}
    \hat{\nabla}_{\phi_i} 
    J_Q\left(\phi_i \right)& = \nabla_{\phi_i}Q_{\phi_i} \left(\boldsymbol S_t, \boldsymbol a_t\right) \Big[ Q_{\phi_i}\left(\boldsymbol S_t, \boldsymbol a_t\right)  \nonumber \\
    &  - \left(r\left(\boldsymbol S_t, \boldsymbol a_t\right) + \gamma Q_{\overline{\phi}_i}\left(\boldsymbol S_{t+1}, \boldsymbol a_{t+1}\right) \right) \Big].\label{equ:grad_Q}
\end{align}
% \color{orange}
% $p$ is the state transition distribution of $\boldsymbol{S}_t$. \color{black}
The loss function of $Q_{\phi_C}\left(\boldsymbol S_t, \boldsymbol a_t\right)$ is 
\begin{align}
J_Q & \left( \phi_C\right)= \mathbb{E}_{\left(\boldsymbol S_t, \boldsymbol a_t\right) \sim \mathcal{B}}\left[\frac{1}{2}\Big(Q_{\phi_C}\left(\boldsymbol S_t, \boldsymbol a_t\right) \right.    \nonumber \\
& -  \left. \left(c\left(\boldsymbol S_t, \boldsymbol a_t\right)+\gamma_C \mathbb{E}_{\boldsymbol a_{t+1} \sim \pi}Q_{\overline{\phi}_C}\left(\boldsymbol S_{t+1}, \boldsymbol a_{t+1}\right)\right)\Big)^2\right], \label{equ:J_Cost}
\end{align}
where $\gamma_C$ is cost discount factor. 
The stochastic gradient to optimize $J_Q\left( \phi_C\right)$ is given by
\begin{align}
\hat{\nabla}_\theta J_Q &\left(\phi_C\right)=\nabla_{\phi_C} Q_{\phi_C}\left(\boldsymbol S_t, \boldsymbol a_t\right) \nonumber  \\
&\times \left[Q_{\phi_C}\left(\boldsymbol S_t, \boldsymbol a_t\right)-\left(c\left(\boldsymbol S_t, \boldsymbol a_t\right)+\gamma_C Q_{\overline{\phi}_C}\left(\boldsymbol S_t, \boldsymbol a_t\right)\right)\right]. \label{equ:grad_Cost}
\end{align}

\subsubsection{Loss function of the policy (actor) network}

%We then introduce the policy function with invariant risk minimization~(IRM) scheme~\cite{IRM_ori} to improve the adaptivity performance of the SAC module.
As the wireless network capacity fluctuates over time, to maintain DT synchronized, the device scheduling policy needs to be continuously adjusted with the changes on the number available RBs (i.e. $M$). The  MTR buffer provides the access to former (fresh and old) device scheduling experiences with different RB constraints, such that the policy function will be trained with invariant risk
minimization (IRM) scheme over common distributions learned from these historical experiences. 
The IRM scheme encourages the policy to learn a state-to-action mapping that is invariant across all possible networking environments, and if the mapping is stable across a large number of different environments, the mapping will be more likely to perform well in the unseen environments~\cite{muli_timescale_rep_CRL}. 
In such context, the loss of IRM actor is given by 
\begin{equation} \label{IRM_define}
J_{\pi}^{IRM}(\theta) = \left\|\nabla_{w \mid w=1} J^{e}_{\pi}(\theta, w \cdot \pi)\right\|^2 , 
\end{equation}
% \begin{equation}
% \min _{\Phi: \mathcal{X} \rightarrow \mathcal{Y}} \sum_{e \in \mathcal{E}_{t r}} R^e(\Phi)+\lambda \cdot\left\|\nabla_{w \mid w=1.0} R^e(w \cdot \Phi)\right\|^2 ,
% \end{equation}
% where $\pi$ is the policy which maps the input states into the probability distribution of action,
% mapping function learned by the actor network, 
within which~$J^{e}_{\pi}(\theta, w \cdot \pi)$ is the loss function of the original SAC policy function parameterized by~$\theta$ under environment~$e$, with $w \cdot \pi$ being a stable policy across multiple environments, and $w$ is a dummy variable~\cite{IRM_ori}. 
The term $\nabla_{w \mid w=1} J^{e}_{\pi}(\theta, w \cdot \pi)$ is the derivative of $J^{e}_{\pi}(\theta, w \cdot \pi)$ at $w=1$, which indicates how much the actor network loss changes as the policy changes, thus the term $\nabla_{w \mid w=1} J^{e}_{\pi}(\theta, w \cdot \pi)$ can be 
used to measure the performance of the dummy policy $w\cdot \pi$ at each environment $e$.
Hence, minimizing $J_{\pi}^{IRM}(\theta)$ can reduce the policy $\pi$ intrinsically change along with new environments, thus mitigating catastrophic forgetting.

% 这里，前四个公式可以写跟FAC一样。从而简短

% \begin{strip}[b]
% \hrule
% \begin{align}\label{equ:J_Q}
% J_Q\left(\phi_i \right)=\mathbb{E}_{\left( \boldsymbol S_t, \boldsymbol a_t\right) \sim \mathcal{B}}   \left[ \frac{1}{2} \left(Q_{\phi_i}\left(\boldsymbol S_t, \boldsymbol a_t\right) - \mathbb{E}_{ \boldsymbol  a_{t+1} \sim \pi} \left(r\left(\boldsymbol S_t, \boldsymbol a_t\right)+ \gamma Q_{\overline{\phi}_i}\left(\boldsymbol S_{t+1}, \boldsymbol  a_{t+1}\right) \right) \right)^2\right], i \in \{1,2\}, 
% \end{align}
% \end{strip}

% \begin{figure*}[b] % hb底部，ht为头部
% \centering % 公式居中
% \vspace*{8pt} % 调整线与公式之间的距离
% \begin{equation}
% \label{formula: bounds on mu}
% 0<\mu \leq \frac{2 E\left\{\operatorname{Re}\left\{f(e[n]) e_a[n]-\lambda \Delta \boldsymbol{w}^H[n] Q(\boldsymbol{w}[n])\right\}\right\}}{\mathrm{E}\left\{\|\boldsymbol{x}[n]\|_2^2|f(e[n])|^2\right\}+\lambda^2 E\left\{\|Q(\boldsymbol{w}[n])\|_2^2\right\}-2 \lambda E\left\{\operatorname{Re}\left\{f(e[n]) Q^H(\boldsymbol{w}[n]) \boldsymbol{x}[n]\right\}\right\}}
% \end{equation}
% \end{figure*}

% gradient of loss function derivative at w=1

% the model that maps the inputs to the outputs (and is a function
% of the model parameters), $R^e$ is the loss function for environment $e$, $w$ is a dummy variable and \color{red} $\lambda$ \color{black} is
% a parameter that balances the importance of the empirical loss (the first term) and the IRM loss (the second term).

The original SAC actor network loss function $J^{SAC}_\pi(\theta)$  is 
\begin{align}
J^{SAC}_\pi(\theta)&=\mathbb{E}_{\boldsymbol S_t \sim \mathcal{B}} \biggl\{ \mathbb{E}_{\boldsymbol a_t \sim \pi }\biggl[\alpha \log \bigl(\pi \left(\boldsymbol a_t \mid \boldsymbol S_t\right)\bigr) \nonumber \\
&-Q_\phi\left(\boldsymbol S_t, \boldsymbol a_t\right)+\lambda_{\zeta}\left(\boldsymbol S_t\right)\bigl(Q_{\phi_C}\left(\boldsymbol S_t, \boldsymbol a_t\right)-M \bigr) \biggr] \biggr\}, \label{equ:J_theta}
\end{align}
where $\alpha$ is the weight of action entropy, $Q_\phi\left(\boldsymbol S_t, \boldsymbol a_t\right)=\min \{Q_{\phi_1}\left(\boldsymbol S_t, \boldsymbol a_t\right), Q_{\phi_2}\left(\boldsymbol S_t, \boldsymbol a_t\right)\}$ is the Q value derived from the two reward value functions, 
$\lambda_{\zeta}\left(\boldsymbol S_t\right)$ is the estimated multiplier value, and
$ Q_{\phi_C}\left(\boldsymbol S_t, \boldsymbol a_t \right)$ is the estimated cost value.

The loss function $J_\pi(\theta)$ of MTR-SAC actor network is given by  
\begin{align}
J_\pi(\theta)& =  J^{SAC}_\pi(\theta) 
 + \lambda_{IRM}\sum_{i=1}^{n_b} \frac{|\mathcal D_i|}{|\mathcal D_{MTR}| }  \nonumber \\
& \times \mathbb{E}_{\boldsymbol S^{i}_{t} \sim \mathcal{D}_i}   \left[ \left\|\nabla_{w \mid w=1} J_\pi^{SAC}\left(\theta, w \cdot \pi , \boldsymbol S^{i}_{t} \right)\right\|^2 \right], \label{equ:loss_IRMactor}
\end{align}
%Based on the MTR buffer's character that each sub-buffer stores the experiences at different timescale history, the loss function  of the actor network $J_\pi(\theta)$ in the designed MTR-SAC is given by~(\ref{equ:loss_IRMactor}),
% \begin{equation}
% \begin{aligned}
% &J_\pi(\theta) = J^{SAC}_\pi(\theta) + \lambda_{IRM} J_{\pi}^{IRM}(\theta) = J^{SAC}_\pi(\theta) +  \\
% &\lambda_{IRM}\sum_{i=1}^{n_b} \frac{|\mathcal D_i|}{|\mathcal D_{MTR}| } \mathbb{E}_{\boldsymbol S^{i}_{t} \sim \mathcal{D}_i} \left[ 
% \left\|\nabla_{w \mid w=1} J_\pi^{SAC}\left(\theta, w \cdot \pi , \boldsymbol S^{i}_{t} \right)\right\|^2 \right] ,
% \end{aligned}
% \end{equation}
% \begin{equation}
% \begin{aligned}
% J_\pi(\theta)& =  J^{SAC}_\pi(\theta) \\
% &+ \lambda_{IRM}\sum_{i=1}^{n_b} \frac{|\mathcal D_i|}{|\mathcal D_{MTR}| }  \\ 
% & \times \mathbb{E}_{\boldsymbol S^{i}_{t} \sim \mathcal{D}_i} \left[ \left\|\nabla_{w \mid w=1} J_\pi^{SAC}\left(\theta, w \cdot \pi , \boldsymbol S^{i}_{t} \right)\right\|^2 \right] ,   \label{equ:loss_IRMactor}
% \end{aligned}
% \end{equation}
% \begin{figure*}[!t]
%   \centering
%   \begin{minipage}{\textwidth}
%   \hrule
%     \begin{align}
%       J_\pi(\theta) = & J^{SAC}_\pi(\theta) + \lambda_{IRM}\sum_{i=1}^{n_b} \frac{|\mathcal D_i|}{|\mathcal D_{MTR}| } \mathbb{E}_{\boldsymbol S^{i}_{t} \sim \mathcal{D}_i} \left[ 
%       \left\|\nabla_{w \mid w=1} J_\pi^{SAC}\left(\theta, w \cdot \pi , \boldsymbol S^{i}_{t} \right)\right\|^2 \right] ,\label{equ:loss_IRMactor}
%     \end{align}
%     \vspace{-0.5cm}
%   \end{minipage}
% \end{figure*}
in which $\lambda_{IRM}$ is the weight coefficient that balances the original actor loss $J^{SAC}_\pi(\theta)$ and the loss of IRM actor $J_{\pi}^{IRM}(\theta)$,
$\mathcal{D}_i$ is the set of experiences in the sub-buffer $i$ with $\left|\mathcal{D}_i\right|$ being the number of experiences,
$\left|\mathcal{D}_{ {MTR
}}\right|$ is the total number of experiences in all the MTR buffer.
Hence, the second term in~(\ref{equ:loss_IRMactor}) indicates the performance of the policy in the environments at different timescales.

To obtain the gradient of the actor network, the policy is represented by a NN transformation form $\boldsymbol a_t = f_\theta \left(\epsilon_t ; \boldsymbol S_t\right)$, with $\epsilon_t$ being an input noise vector sampled from a fixed Gaussian distribution~\cite{Soft_Actor_Critic}, since the target of $J^{SAC}_\pi(\theta)$ is the Q-function learned by differentiable NNs.
Take the NN transformation into the policy loss and then, the estimated policy gradient of the original SAC $J^{SAC}_\pi(\theta)$ is
\begin{align}
\hat{\nabla}_\theta J^{SAC}_\pi(\theta) & = \nabla_\theta \alpha \log \left(\pi \left(\boldsymbol a_t \mid \boldsymbol S_t\right)\right) \nonumber \\
& + \Big[ \nabla_{\boldsymbol a_t} \alpha \log \left(\pi \left(\boldsymbol a_t \mid \boldsymbol S_t\right)\right) \nonumber \\
&  - \nabla_{\boldsymbol a_t} \Big( Q_\phi \left( \boldsymbol S_t, \boldsymbol a_t \right) + \lambda_{\zeta} \left( \boldsymbol S_t \right) Q_{\phi_C} \left( \boldsymbol S_t, \boldsymbol a_t \right) \Big) \Big] \nonumber \\
& \times \nabla_\theta f_\theta \left( \epsilon_t ; \boldsymbol S_t \right), \label{equ:grad_theta}
\end{align}
% \begin{figure*}[!b]
%   \centering
%   \begin{minipage}{\textwidth}
%   \vspace{-0.3cm}
%   \hrule
%     \begin{align}
%       \hat{\nabla}_\theta J^{SAC}_\pi(\theta)= & \nabla_\theta \alpha \log \left(\pi \left(\boldsymbol a_t \mid \boldsymbol S_t\right)\right)+ \left[\nabla_{\boldsymbol a_t} \alpha \log \left(\pi \left(\boldsymbol a_t \mid \boldsymbol S_t\right)\right)-\nabla_{\boldsymbol a_t}\left(Q_\phi\left(\boldsymbol S_t, \boldsymbol a_t\right)+   \lambda_{\zeta}\left(\boldsymbol S_t\right) Q_{\phi_C}\left(\boldsymbol S_t, \boldsymbol a_t\right) \right)\right] \nabla_\theta f_\theta\left(\epsilon_t ; \boldsymbol S_t\right),    
%       \label{equ:grad_theta}
%     \end{align}
%     \hrule
%   \end{minipage}
% \end{figure*}
where $M$ is neglected since it is irrelevant to  $\theta$. 
Given $\hat{\nabla}_\theta J^{SAC}_\pi(\theta)$, the gradient of the MTR-SAC method is
\begin{align}
\hat{\nabla}_\theta J_\pi(\theta) & = \hat{\nabla}_\theta J^{SAC}_\pi(\theta) + \lambda_{IRM}\sum_{i=1}^{n_b} \frac{|\mathcal D_i|}{|\mathcal D_{MTR}| } \nonumber \\
& \times \hat{\nabla}_\theta \left[ \left\|\nabla_{w \mid w=1} J_\pi^{SAC}\left(\theta, w \cdot \pi, \boldsymbol S^{i}_{t} \right)\right\|^2 \right]. \label{equ:grad_IRMactor}
\end{align}

% \begin{figure*}[!t]
%   \centering
%   \begin{minipage}{\textwidth}
%     \vspace{-0.3cm}
%     \begin{align}
%     \hat{\nabla}_\theta J_\pi(\theta) = & \hat{\nabla}_\theta J^{SAC}_\pi(\theta) + \lambda_{IRM}\sum_{i=1}^{n_b} \frac{|\mathcal D_i|}{|\mathcal D_{MTR}| } 
%     \hat{\nabla}_\theta \left[ 
%     \left\|\nabla_{w \mid w=1} J_\pi^{SAC}\left(\theta, w \cdot \pi, \boldsymbol S^{i}_{t} \right)\right\|^2 \right].\label{equ:grad_IRMactor}
%     \end{align}
%     \hrule
%     \vspace{-0.5cm}
%   \end{minipage}
% \end{figure*}

% \begin{equation}
%     \hat{\nabla}_\theta J_\pi(\theta) = \hat{\nabla}_\theta J^{SAC}_\pi(\theta) + \lambda_{IRM}\sum_{i=1}^{n_b} \frac{|\mathcal D_i|}{|\mathcal D_{MTR}| } \mathbb{E}_{\boldsymbol S_{t} \sim \mathcal{D}_i} \left[ 
% \left\|\nabla_{w \mid w=1} J_\pi^{SAC}\left(\theta, w \cdot \pi \right)\right\|^2 \right]
% \end{equation}

\subsubsection{Loss function of the multiplier network}
The loss function of 
multiplier network is given by
% is the same as (\ref{equ:J_theta}), which is simplified by
\begin{equation}
J_\lambda(\zeta)=\mathbb{E}_{\boldsymbol S_t \sim \mathcal{B}}\left\{\mathbb{E}_{\boldsymbol a_t \sim \pi }\left[\lambda_{\zeta}\left(\boldsymbol S_t\right)\left(Q_{\phi_C}\left(\boldsymbol S_t, \boldsymbol a_t\right)-M\right)\right]\right\},
\end{equation}
where the entropy and reward value function term are neglected since the irrelevance with multiplier $\lambda$~\cite{thn_conf_v_SCaware_DT}. The stochastic gradient of $\lambda$ can be given by
\begin{equation}
\hat{\nabla} J_\lambda(\zeta)=\left(Q_{\phi_C}\left(\boldsymbol S_t, \boldsymbol a_t\right)-M\right) \nabla_{\zeta} \lambda_{\zeta}\left(\boldsymbol S_t\right).
\end{equation}
%The specific training procedure of the proposed MTR-SAC method is summarized in Algorithm~\ref{alg:update}.

\subsubsection{Training process of the MTR-SAC algorithm} As summarized in Algorithm~\ref{alg:update}, starting with initial actor $\theta$, critic $\phi$, and multiplier networks $\zeta$, the algorithm interacts with the physical networks and uses the achieved state transitions $( \boldsymbol S_t, \boldsymbol a_t, \boldsymbol S_{t+1} )$ and
regressed values
$r(\boldsymbol{S}_t, \boldsymbol a_t)$ and $ c(\boldsymbol{S}_t, \boldsymbol a_t)$ to calculate the loss functions (\ref{equ:J_Q}), (\ref{equ:J_Cost}), and (\ref{equ:J_theta}). It then uses a mini-batch training
mechanism to update value networks. 
Then, based on the sampled interaction experiences in the MTR buffer, it updates the policy network every $m_\pi$ gradient steps, and updates the multiplier network every $m_\lambda$ gradient steps. 
Such trial, error, then update procedure will be repeated, until the convergence is reached.
{\color{black}In Algorithm~\ref{alg:update},  $\beta_{Q}$, $\beta_{\pi}$,$\beta_{\alpha}$, $\beta_{\lambda}$, and $\rho$ are learning rates of corresponding NNs.}

\subsection{Convergence, Implementation, and Complexity Analysis} 
In this section, we analyze the convergence, implementation and complexity of the proposed MTR-SAC based CRL device scheduling method.  

\begin{algorithm}[t]
	\renewcommand{\algorithmicrequire}{\textbf{Input:}}
	\renewcommand{\algorithmicensure}{\textbf{Output:}}
	\caption{\small Training process of the MTR-SAC algorithm}
	\label{alg:update}
	% \small
	\begin{algorithmic}[1]
% 		\REQUIRE latent dimension $K$, $G$, target predicate $p$
% 		\ENSURE $U^{p}$, $V^{p}$, $b^{p}$
		\STATE \textbf{Input:} $\phi_1, \phi_2, \phi_C, \theta, \zeta$
  		\STATE \textbf{Initialize:} $\mathcal{B} \leftarrow \emptyset$, 
    % $\bar{\diamondsuit} \leftarrow \diamondsuit$, 
    %  $\bar{\phi_2} \leftarrow \phi_2$, and $\bar{\phi_C} \leftarrow \phi_C$
    $\bar{\diamondsuit} \leftarrow \diamondsuit$ for $\diamondsuit \in\left\{\phi_1, \phi_2, \phi_C, \theta\right\}$
            \STATE \textbf{for} each episode \textbf{do}
            \STATE \quad \textbf{for} each environment interaction time slot \textbf{do}
            \STATE \quad \quad 
            BS obtain $\boldsymbol{S}_t$,  $\boldsymbol a_t, \boldsymbol{S}_{t+1}, r(\boldsymbol{S}_t, \boldsymbol a_t),  c(\boldsymbol{S}_t, \boldsymbol a_t)$ and \\
            \quad \quad fill the  MTR buffer by \\
            \quad \quad $\boldsymbol a_t \sim \pi_\theta\left(\boldsymbol a_t \mid \boldsymbol S_t\right), \boldsymbol S_{t+1} \sim p\left(\boldsymbol S_{t+1} \mid \boldsymbol S_t, \boldsymbol a_t\right), $ \\
            % $\hfill\triangleright$ Sample transitions
            \quad \quad $\mathcal{B} \leftarrow \mathcal{B} \cup\left\{\left(\boldsymbol S_t, \boldsymbol a_t, r\left(\boldsymbol S_t, \boldsymbol a_t\right), c\left(\boldsymbol S_t, \boldsymbol a_t\right), \boldsymbol S_{t+1}\right)\right\}$  
            % $\hfill\triangleright$ Store the transition in the replay buffer
            \STATE \quad \textbf{end for}
            \STATE \qquad \textbf{for} each gradient step \textbf{do}
            \STATE \qquad \quad update value networks by \\
            \qquad \quad $\phi_i \leftarrow \phi-\beta_Q \hat{\nabla}_{\phi_i} J_Q\left(\phi_i\right)$ for $i \in\{1,2, C\}$
            \STATE \qquad \textbf{if}  gradient steps mod $m_\pi=0$ \textbf{then}
            \STATE \qquad \quad update the policy network by \\
            \qquad \quad  $\theta \leftarrow \theta-\beta_\pi \hat{\nabla}_\theta J_\pi(\theta) $
            \STATE 
            \qquad \quad adjust the temperature by \\
            \qquad \quad $\alpha \leftarrow \alpha-\beta_\alpha  \hat{\nabla}_\alpha J_\alpha(\alpha)$
            \STATE \qquad \textbf{end if}
            \STATE \qquad \textbf{if} gradient steps mod $m_\lambda=0$  \textbf{then}
            \STATE \qquad \quad update the multiplier network by \\
            \qquad \quad $\zeta \leftarrow \zeta+\beta_\lambda \nabla_{\zeta} J_\lambda(\zeta)$
            \STATE \qquad \textbf{end if}
            \STATE \qquad  update the target networks by\\
            \qquad $\bar{\diamondsuit} \leftarrow \rho \diamondsuit+(1-\rho) \bar{\diamondsuit}$ \ for $\diamondsuit \in\left\{\phi_1, \phi_2, \phi_C, \theta\right\}$
            \STATE \quad \textbf{end for}
            \STATE  \textbf{end for}
            \STATE  \textbf{Output:} $\phi_1, \phi_2, \phi_C, \theta, \zeta$
            \end{algorithmic}  
    	\vspace{-0.cm}
\end{algorithm}

\textit{1) Convergence Analysis:}
For the CMDP in~(\ref{equ:p_min_aoii}) in dynamic environments, we cannot give the optimal solution that the proposed algorithm reaches. 
Next, using the result of~\cite[Theorem 1]{Soft_Actor_Critic}, we can prove that the proposed algorithm will not diverge and can converge to a fixed solution {\color{black} of problem (\ref{equ:prob_2})} in a static environment, as shown in the following

% Next, using the result of Theorem 1 in \cite{Soft_Actor_Critic}, we can prove that
% the proposedMTR-SAC method can converge as shown in the following

\textbf{Lemma~1:} 
\textit{The proposed MTR-SAC method is guaranteed
to converge if the following conditions are satisfied
1) reward value function $Q(\boldsymbol S_t, \boldsymbol a_t )$ is bounded, and 
2) $Q^{{\pi}_{\text{new}}}  (\boldsymbol S_t, \boldsymbol a_t ) \geq Q^{{\pi}_{\text{old}}}  (\boldsymbol S_t, \boldsymbol a_t )$ holds for feasible $\boldsymbol S_t$ and  $\boldsymbol a_t$ with ${\pi}_{\text{new}}$ being the updated new policy from the old policy $\boldsymbol{\pi}_{\text{old}}$ in each iteration.
}

% hence $Q(\boldsymbol S_t, \boldsymbol a_t )$ is bounded

\textit{Proof.} 
The proposed MTR-SAC satisfies the two conditions for the following reasons.
First, since the actor network is differentiable NN~\cite{Soft_Actor_Critic}, the value of  $J_{\pi}^{IRM}(\theta)$ is bounded, and the finite dimension of $\boldsymbol a_t$ guarantees the action entropy is bounded.
Hence the $Q(\boldsymbol S_t, \boldsymbol a_t )$ is bounded and  satisfies the condition 1).

Second, the case $J_{{\pi}_{\text{old}}} ( \pi_{\text{new}}( \cdot|\boldsymbol S_t ) ) 
\leq J_{{\pi}_{\text{old}}} ( \pi_{\text{old}}( \cdot|\boldsymbol S_t ) )$ always holds in the static environment  
because we can always choose ${\pi}_{\text{new}} = {\pi}_{\text{old}} \in \Pi$ for the invariant distribution of $\mathcal{D}_i$.
And because the second term in~(\ref{equ:J_theta}) is only relative to the parameter $\theta$, according  to the Lemma 2 in~\cite{Soft_Actor_Critic}, the above inequality can be reduced to 
\begin{equation}\label{equ:reduced}
\begin{aligned}
&\mathbb{E}_{\boldsymbol{a}_t \sim \pi_{\text {new }}}\left[\log \pi_{\text {new }}\left(\boldsymbol{a}_t \mid \boldsymbol{S}_t\right)-Q^{\pi_{\text {old }}}\left(\boldsymbol{S}_t, \boldsymbol{a}_t\right)\right] \\
&\leq  
\mathbb{E}_{\boldsymbol{a}_t \sim \pi_{\text {old }}}\left[\log \pi_{\text {old }}\left(\boldsymbol{a}_t \mid \boldsymbol{S}_t\right)-Q^{\pi_{\text {old }}}\left(\boldsymbol{S}_t, \boldsymbol{a}_t\right)\right].
\end{aligned}
\end{equation}
Given~(\ref{equ:reduced}), consider the soft Bellman equation, it can be obtained that $Q^{{\pi}_{\text{old}}}  (\boldsymbol S_t, \boldsymbol a_t ) \leq Q^{{\pi}_{\text{new}}}  (\boldsymbol S_t, \boldsymbol a_t )$, which satisfies the condition 2).
$\hfill\blacksquare$
\\

% The convergence of  the proposed MTR-SAC method can be given by 

\begin{comment}

分析 SAC 是因为动作空间有限+bellman才收敛的，现在看看bellman 是否继续满足

\textbf{Lemma 1 
% (Soft Policy Evaluation Convergence):
}
\textit{Consider the soft Bellman backup operator $\mathcal{T}^\pi$ and a mapping
$Q^0: \mathcal{S} \times \mathcal{A} \rightarrow \mathbb{R}$ with $|\mathcal{A}|<\infty$, and define  $Q^{k+1}=\mathcal{T}^\pi Q^k$.
Then the sequence $Q^k$ will converge to the soft Q-value of $\pi$ as $k \rightarrow \infty$.
}

\textit{Proof.} Define the entropy augmented reward as $r_\pi\left(\boldsymbol{S}_t, \boldsymbol{a}_t\right) \triangleq r\left(\boldsymbol{S}_t, \boldsymbol{a}_t\right)+\mathbb{E}_{\boldsymbol{S}_{t+1} \sim p}\left[\mathcal{H}\left(\pi\left(\cdot \mid \mathbf{S}_{t+1}\right)\right)\right]$ and rewrite the update
rule as
\begin{equation}
Q\left(\boldsymbol{S}_t, \boldsymbol{a}_t\right) \leftarrow r_\pi\left(\boldsymbol{S}_t, \boldsymbol{a}_t\right)+\gamma \mathbb{E}_{\boldsymbol{S}_{t+1} \sim p, \boldsymbol{a}_{t+1} \sim \pi}\left[Q\left(\boldsymbol{S}_{t+1}, \boldsymbol{a}_{t+1}\right)\right]
\end{equation}
and apply the standard convergence results for policy evaluation~\cite{RL_book}. The assumption $|\mathcal{A}|<\infty$ is required to guarantee that the entropy augmented reward is bounded. 

Since the actor network is differentiable neural network~\cite{Soft_Actor_Critic}, the value of  $J_{\pi}^{IRM}(\theta)$ is bounded.

\end{comment}

\textit{2) Implementation Analysis:}
Next, we explain the implementation of the proposed MTR-SAC method.
The devices use the feedback message from the BS to obtain the virtual state $\hat{X}_t$, and then calculate the mismatch value $Z_{n,t}$. 
% In this way, devices can deploy the efficient data sending scheme which only sends mismatch-aware sensing data to verify the mismatch.  
% Before sending data to BS, the devices must request RB for RB allocation which can deduce the mismatch of devices. 
% The mismatch value $Z_{n,t}$ can be 
The BS can only obtain the mismatch value $Z_{n,t}$ by scheduling device $n$, as $Z_{n,t}$ is transmitted to the BS along with the sensing data packets. 
Since the BS can not collect real-time $Z_{n,t}$ of all devices due to limited RBs, the BS must record the last received mismatch value $Z_{n,g_{n,t}^{\mathrm{m}}}$ as an estimation of $Z_{n,t}$.
% $Y_{n,t} = Z_{n,g_{n,t}^{\mathrm{m}}}$ being the latest received mismatch value
Meanwhile, the BS records the last received packet from the devices, such that it can calculate~$\varPhi_{n,t}(u_{n,t})$ as well as the packet reception indicator~$\gamma_{n,t}$.
When the resource budget changes, the BS can obtain the total number of available RBs $M$ for device scheduling.
In other words, the BS is capable of accessing the necessary information and computation capability to train and deploy the proposed algorithm.
%The device scheduling method is practical to be implemented in the BS.

\textit{3) Complexity Analysis:}
The time complexity of training and inferring  the algorithm is mainly determined by the size of the NN weights~\cite{thn_FedAudio}. 
Let $N_l$ be the total number of layers in each NN in SAC module, $h_i$ be the number of neurons in layer $l_i$, $h_{in}$ be the dimension of the input, and $h_{out}$ be the dimension of the output.
Thus, the time complexity of inferring the proposed algorithm can be calculated by $\mathcal{O}\left( h_{in} h_1 +  \sum_{i=1}^{N_l -1} h_{i} h_{i+1} + h_{out} h_{N_l} \right)$.
In the training stage, the MTR-SAC method additionally runs the MTR buffer $J_{\pi}^{IRM}(\theta)$.
The operation of MTR buffer includes pushing in and popping out one experience, thus having the time-complexity of $\mathcal{O}\left(n_b\right)$.
The time complexity of the training policy network is  $\mathcal{O}\left( h_{in} h_1 +  \sum_{i=1}^{N_l -1} h_{i} h_{i+1} + h_{out} h_{N_l} + n_b \right)$.

\color{black}

\begin{table}[t]
  \centering
  \renewcommand{\arraystretch}{1.3}
  \caption{Simulation Parameters}
  \label{tab:para}
  \begin{tabular}{|c|c|c|c|}
    \hline
    \textbf{Parameter} & \textbf{Value} & \textbf{Parameter} & \textbf{Value} \\
    \hline
    $N$ & 20 & $M$ & 18 \\
    \hline
    $W$ & 180kHz & $L_n$ & 250B \\
    \hline
    $P_{n}^{\mathrm{u}}$ & 0.5W & $m$ & 0.023dB \\
    \hline
    $N_0$ & -175 dBm/Hz & $\xi^{\mathrm{th}}$ & $1 \times10^{-2}$ \\
    \hline
    $\beta_{Q}$, $\beta_{\pi}$ & $3\times10^{-4}$ & $\beta_{\alpha}$, $\beta_{\lambda}$ & $1 \times10^{-5}$ \\
    \hline
    $\rho$ & $5\times10^{-3}$ & $N_l$ & 3 \\
    \hline
    \textcolor{black}{$h_{in}, h_{out}$} & \textcolor{black}{$4N, N$} & $h_1,h_2,h_3$ & 256 \\
    \hline
    $\beta_{{mtr}}$ & 0.8 & $\lambda_{IRM}$ & $1 \times10^{-2}$ \\
    \hline
    $n_b$ & 4 & $N_b$ & $5\times10^{3}$ \\
    \hline
    $m_\pi$ & 2 & $m_\lambda$ & $12$ \\
    \hline
  \end{tabular}
  \vspace{-0.3cm}
\end{table}

\section{Simulation Results and Discussion}

In our simulation, we consider a wireless network with 20 wireless devices, including 8 thermometers, 8 hygrometers, and 4 positioning sensors. 
The transmission priority weights $\boldsymbol{w}$ of the thermometers, hygrometers, and positioning are $\frac{3}{20}, \frac{1}{10}, \frac{1}{20}$, respectively, and the unit cost of RB $\boldsymbol{b}$ for the data transmission at each type of device is $1, 1, 5$, respectively.
In this setting, scheduling all devices at each time slot consumes $\sum ^{N}_{n=1} b_n = 36$ RBs while only $M$~($M < 36$) RBs are available for DT synchronization.
The proposed algorithm is trained and inferred with combined real-world sensing datasets from  Intel Berkeley research lab sensor data~\cite{data_intel_lab} and indoor positioning data~\cite{trajectory_data}. 
% 10,000 data records are used for training the algorithm, 2000 for validating, and 2000 records for testing, respectively.
Other parameters are listed in Table~\ref{tab:para}.
{\color{black}In our simulations, the trained model requires approximately 5 MB of storage and performs each inference with 0.1 ms on an Intel Core i7-13700KF CPU.}

\begin{figure}[t]
    \centering
    \vspace{-0.cm}
    \includegraphics[width=1.\linewidth,]{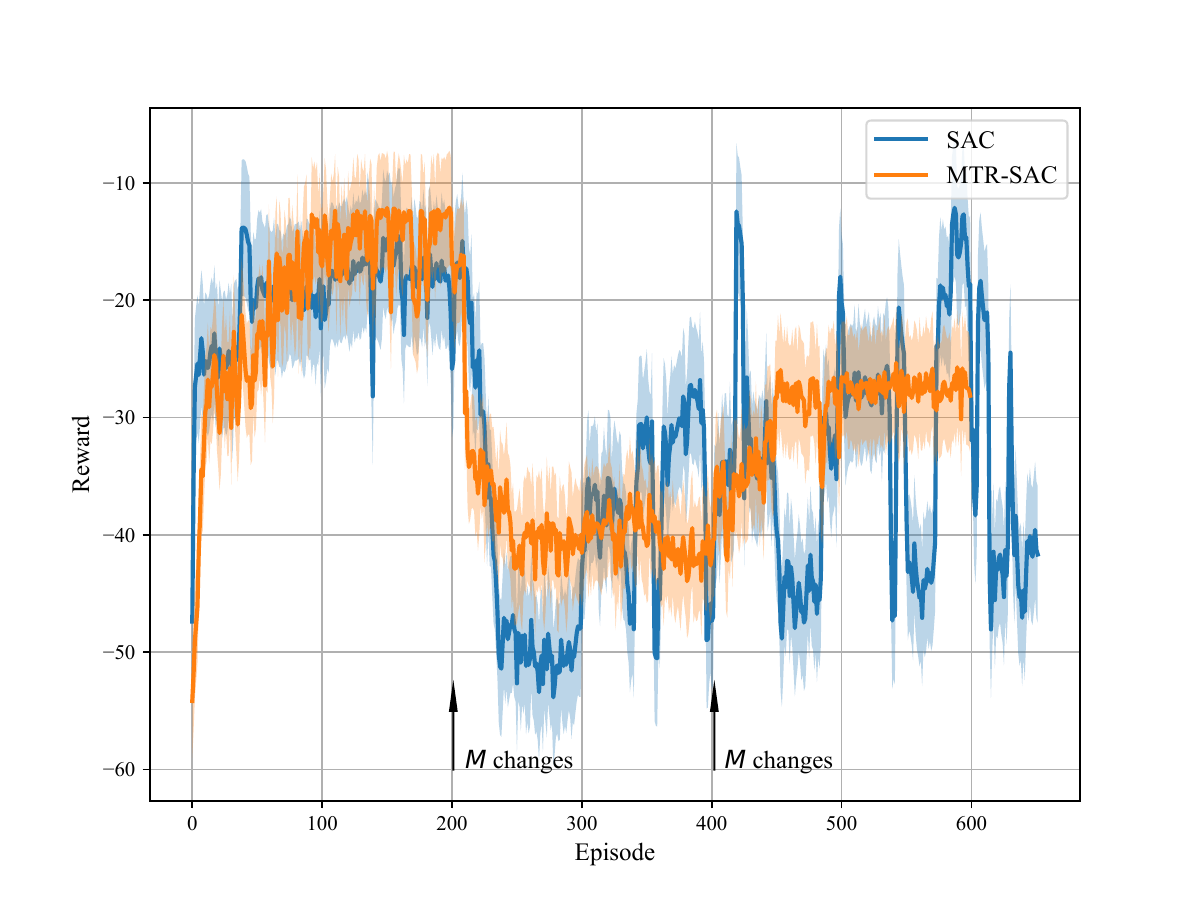}
    \setlength{\abovecaptionskip}{-0.5cm}
    \caption{Convergence of the MTR-SAC method with variable numbers of RBs. The number of available RBs changes from $30$, to $10$, and then $26$.}\label{fig:res_converg}
    % \vspace{-0.3cm}
\end{figure}

\begin{figure}[t]
    \centering
    \includegraphics[width=1.\linewidth,trim=0 0 0 0, clip]{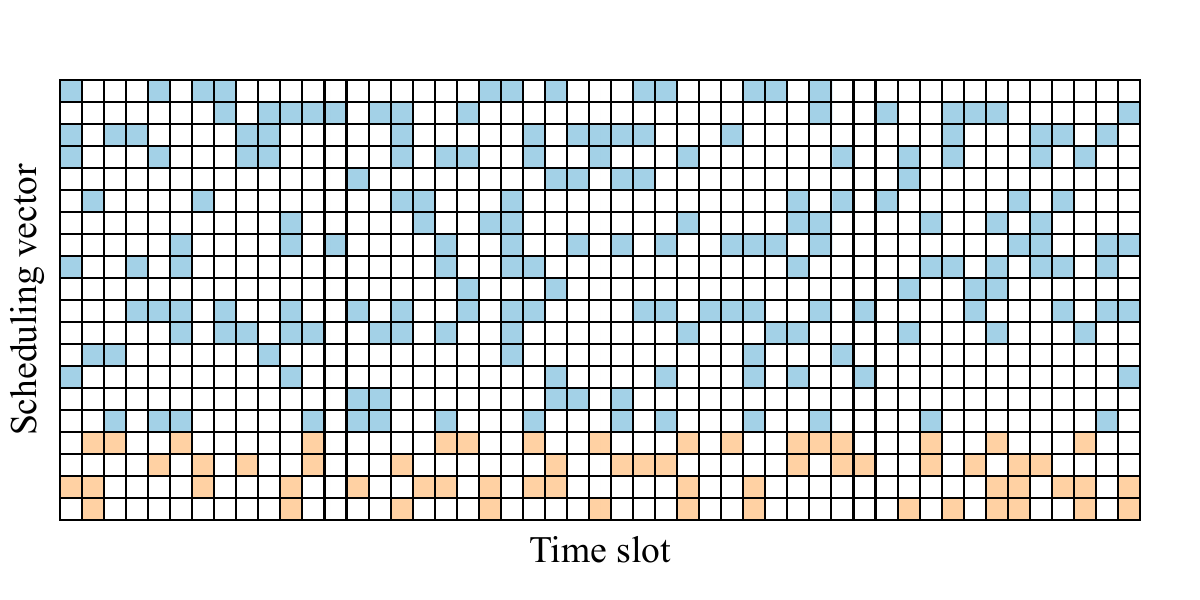}
    \setlength{\abovecaptionskip}{-0.3cm}
    \caption{Device scheduling vectors of the proposed CRL algorithm. $M=18$. The white blocks indicate $u_{n,t}=0$, and colored blocks indicate $u_{n,t}=1$. In each scheduling vector, each black block consumes 1 RB per transmission, and each orange block consumes 5 RBs per transmission.
    }\label{fig:action_show}
    \vspace{-0.3cm}
\end{figure}

The proposed algorithm is compared with three baselines:
a) Standard SAC method that schedules devices without the assistance of MTR buffer,
b) Polling method that schedules each device by turns, and 
c) DP based method that schedules each device with fixed time slot intervals solved by DP~\cite{Aoii_TWC}.
The comparison between SAC and the proposed CRL method can justify how the {\color{black} continual learning (CL)} mechanism improves convergence performance of the scheduling algorithm within dynamic wireless network capacity. 
The comparison between the proposed CRL algorithm and the polling method will justify how adaptive scheduling promotes high-accuracy estimation within DT, while the comparison with the DP method demonstrates that the perception of channel conditions in the proposed CRL algorithm further enhances the DT mapping accuracy.

\begin{figure*}[t]
    \centering
    % 第一行
    \subfigure[Estimated temperature]{\includegraphics[width=0.45\linewidth, height=0.31\linewidth, trim=30 12 45 40, clip]{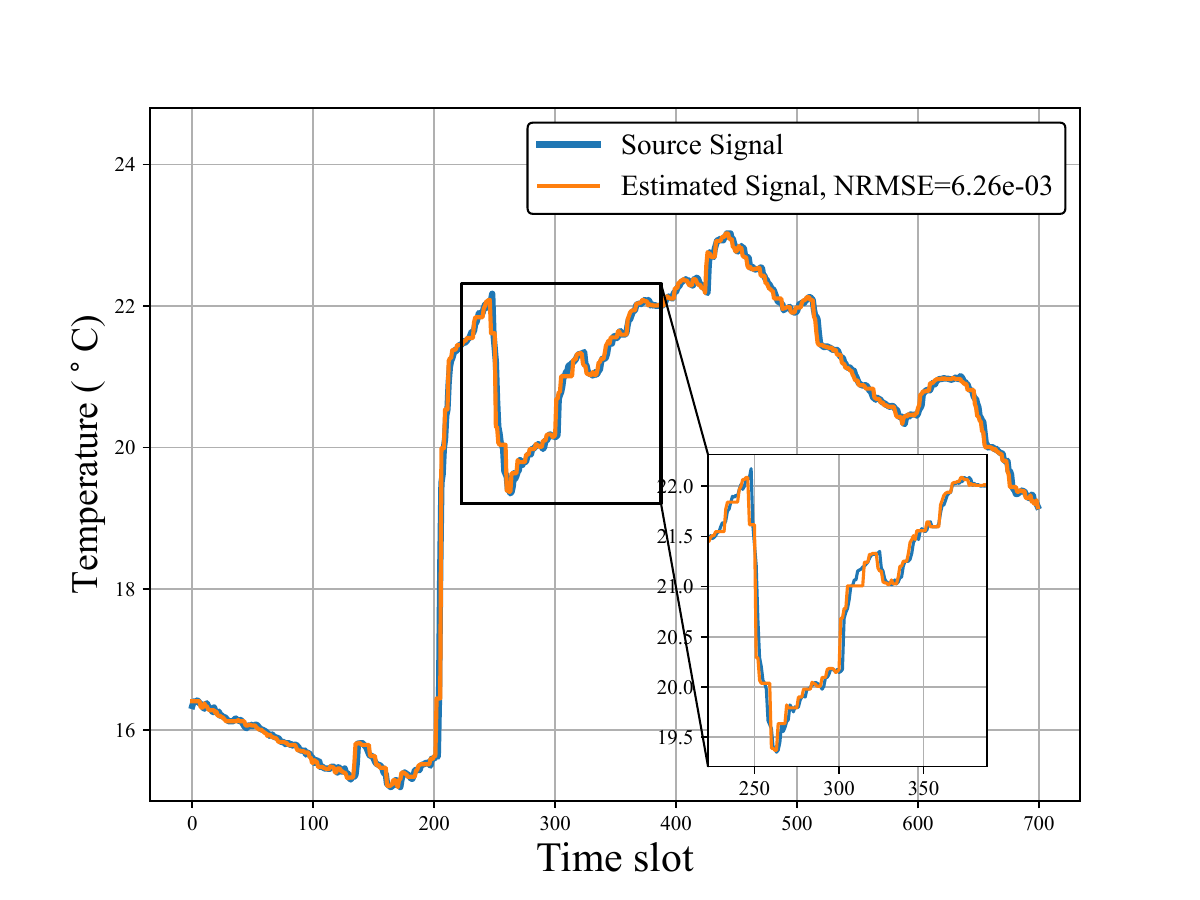}\label{fig:image1}}
    \subfigure[Estimated humidity]{\includegraphics[width=0.45\linewidth, height=0.31\linewidth, trim=30 12 45 40, clip]{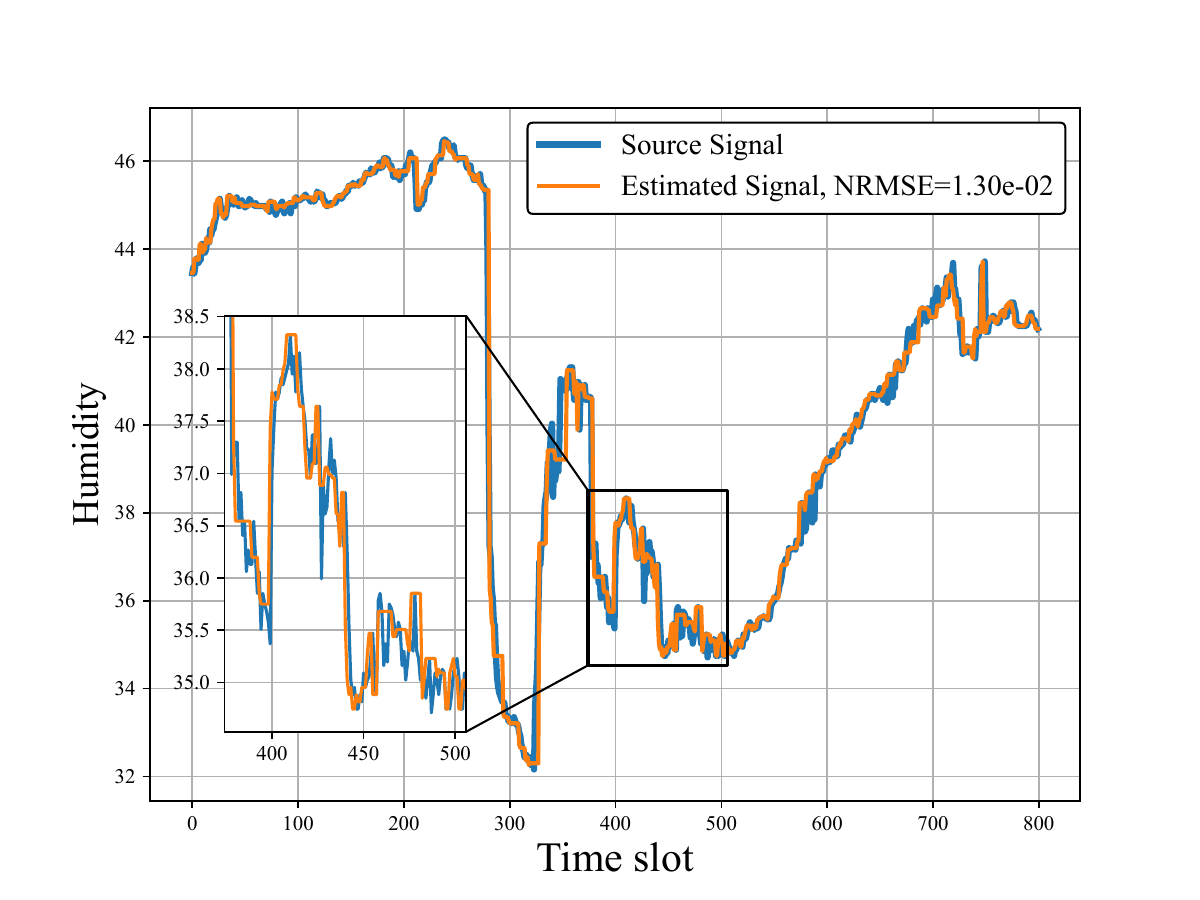}\label{fig:image2}}
    
    \vspace{0.cm}
    % 第二行
    \subfigure[Estimated one-dimensional location]{\includegraphics[width=0.45\linewidth, height=0.31\linewidth, trim=30 12 45 40, clip]{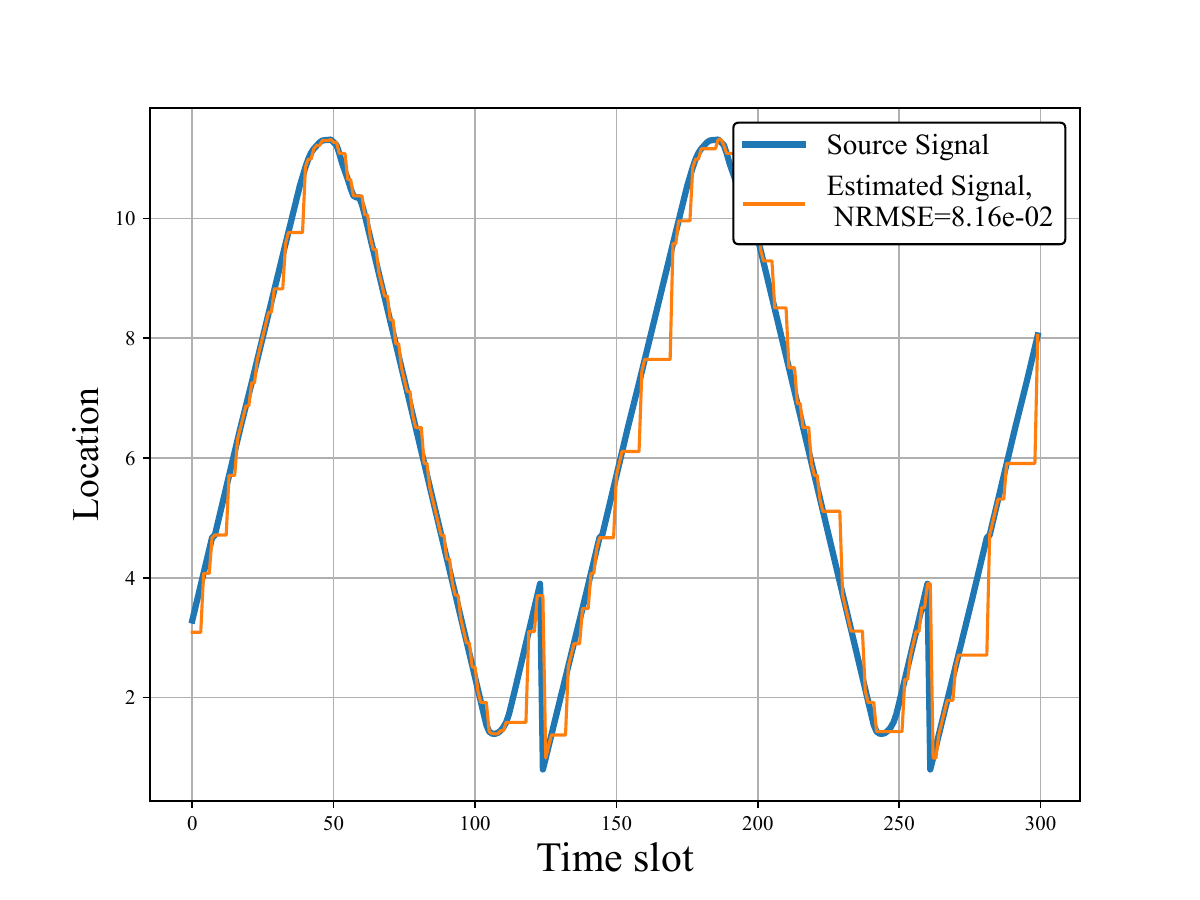}\label{fig:image3}}
    \subfigure[Estimated two-dimensional trajectory]{\includegraphics[width=0.45\linewidth, height=0.31\linewidth, trim=30 12 45 40, clip]{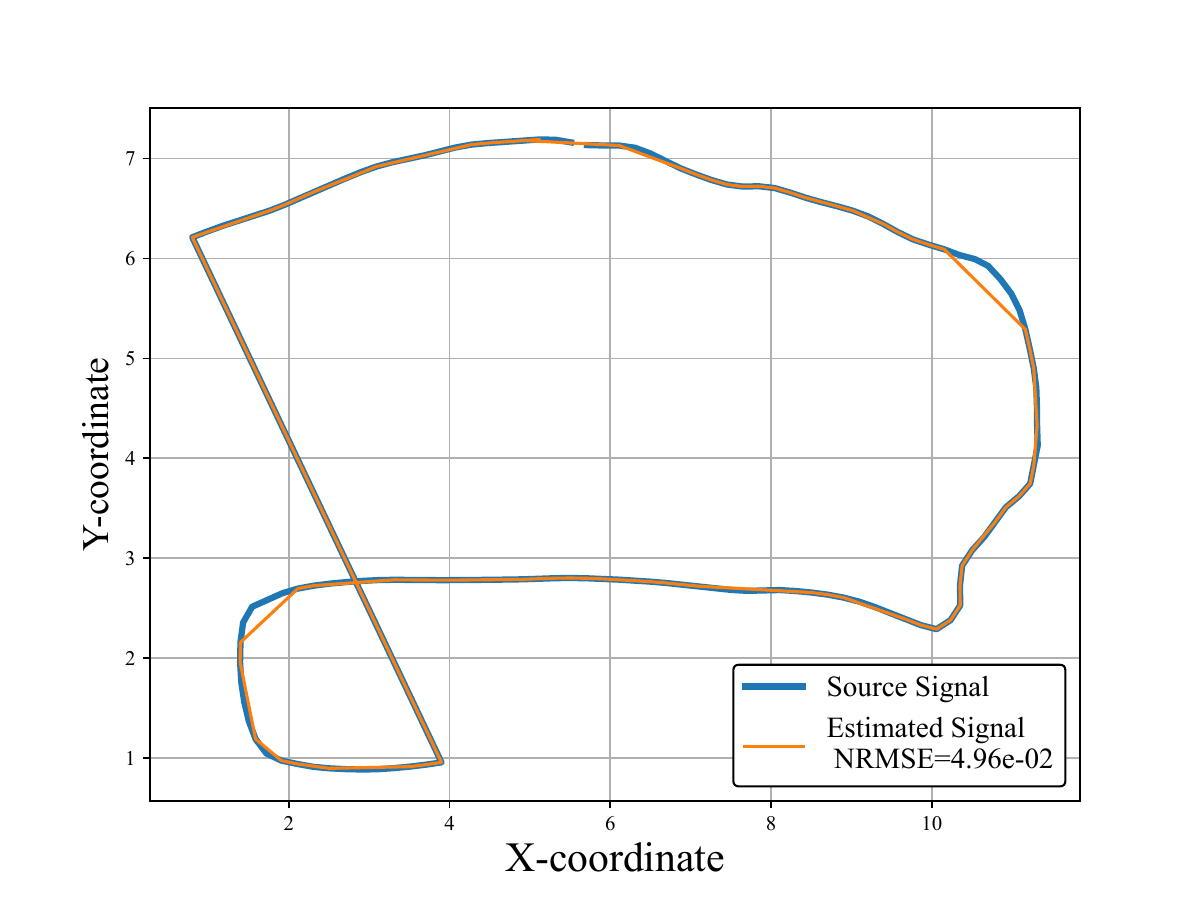}\label{fig:image4}}
    
    \caption{Estimated virtual state signals and the physical state signals.}
    \label{fig:res0}
    \vspace{-0.5cm}
\end{figure*}

Fig.~\ref{fig:res_converg} shows how the proposed MTR-SAC algorithm and standard SAC method converge with changing wireless resources. 
In the simulation, the number of available RBs $M$ drops abruptly from 30 to 10 at the 200th episode, then increases to 26 at the 400th episode.
From Fig.~\ref{fig:res_converg} we see that, the rewards of both the methods decrease and increase with the available RB resources as expected.
This stems from the fact that the introduced multiplier function can adjust multiplier $\lambda$, thereby enabling the agent to adapt to the RB constraint changes. 
% We also observe that, the proposed MTR-SAC method 
% and finally converge as $M$ changes along with the episode. 
Fig.~\ref{fig:res_converg} also shows that the reward of the SAC method fluctuates dramatically after $M$ changes, while that of the MTR-SAC method remain relatively staple. 
This is because the adopted IRM loss in the MTR scheme can restrain drastic updates of the policy.
From Fig.~\ref{fig:res_converg}, we can also see that, after $M$ changes at the 200th and the 400th episodes, the MTR-SAC method quickly converges while the standard SAC method stays unconverged. 
This is because the standard SAC method suffers from catastrophic forgetting of old experiences after $M$ changes, 
while the MTR buffer scheme and IRM loss refine the common knowledge across different constraints, thereby achieving quick adaptation to new constraints. 
{\color{black} Fig.~\ref{fig:res_converg} also shows that, the proposed MTR-SAC method converges slightly slower than SAC during the initial  convergence phase while achieving faster convergence in subsequent phases. 
This is because the adopted IRM loss in MTR moderates drastic changes in the policy network thus facilitating the learning of common knowledge across diverse constraints.
Consequently, the MTR-SAC exhibits slightly slower initial convergence, followed by quicker convergence in dynamics with the learned common knowledge.}
% in initial convergence the MTR-SAC is slightly slower, followed by a faster convergence thereafter with dynamics.
% resulting in slightly slow convergence speed at the beginning and faster speed then.
Thus, the simulation results shown in
Fig.~\ref{fig:res_converg} can justify the improvement in few shot performance of SAC algorithm with the integrated MTR scheme.

% \begin{figure*}[t]
%   \centering
%   \includegraphics[width=0.4\linewidth]{figs/J_res/res0_orisign_vs_estsign_temperature.pdf}
%   \hfill
%   \includegraphics[width=0.4\linewidth]{figs/J_res/res0_oris_vs_ests_humidity_.pdf}
%   \hfill
%   \includegraphics[width=0.4\linewidth]{figs/J_res/res0_oris_vs_ests_periodic_.pdf}
%   \hfill
%   \includegraphics[width=0.4\linewidth]{figs/J_res/res0_orisign_vs_estsign_trajectory_available.pdf}
%   \caption{Sampling result visualization.}
%   \label{fig:res0}
% \end{figure*}

Fig.~\ref{fig:action_show} shows a visualization of device scheduling vectors in a fragment of successive time slots. 
% In this simulation, $M$ is set as 18.
In Fig.~\ref{fig:action_show} we see that, the proposed CRL algorithm can schedule the on-demand devices with successive time slots, thus guaranteeing the sensed physical state changes are transmitted with enough scheduled RBs.
% This phenomenon stems from the fact that the adopted mismatch loss functions extract the significant state changes, and these changes can be detected by the proposed CRL algorithm, which then schedules transmission for a period accordingly.
This phenomenon stems from the fact that the adopted mismatch loss functions extract significant state changes, which can be detected by the proposed CRL algorithm. The CRL algorithm then schedules RBs for an appropriate period accordingly.
Fig.~\ref{fig:action_show} also shows that each device can be scheduled once within an interval of several time slots, thereby maintaining fairness in device scheduling.
This is because the CRL algorithm uses action entropy maximization loss in (\ref{equ:J_theta}) to generate stochastic actions.   

% proposed policy can perceive the 
% Fig.~\ref{fig:action_show} shows that the proposed method can perceive the state variation and 

\begin{figure}[t]
    \centering
    % \vspace{-0.5cm}
    \includegraphics[width=0.98\linewidth,]{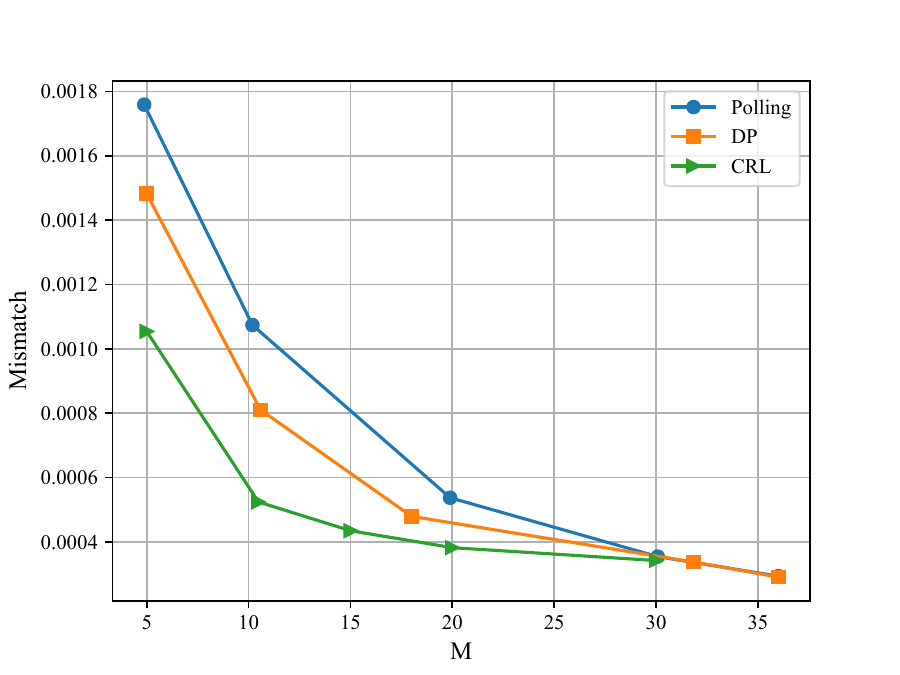}
    \setlength{\abovecaptionskip}{-0.cm}
    \caption{ Weighted mismatch as the number of RBs $M$ increases. In the simulation, all sensing data $X_{n,t}$ is normalized for the convenience of comparison.  
}\label{fig:res1}
\vspace{-0.2cm}
\end{figure}

\begin{figure}[t]
    \centering
    \includegraphics[width=1.\linewidth,]{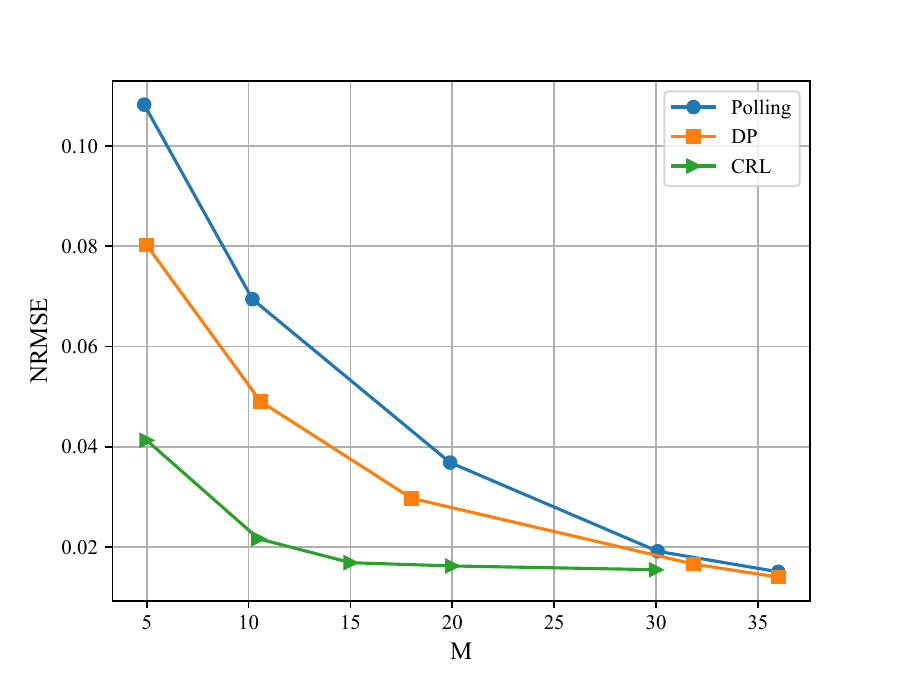}
    \setlength{\abovecaptionskip}{-0.0cm}
    \caption{NRMSE of devices as number of RBs $M$ increases.}\label{fig:res4}
    \vspace{-0.cm}
\end{figure}

Fig.~\ref{fig:res0} shows 4 examples of the estimated virtual states and the corresponding estimation NRMSE. 
Notice that the dramatic changes in the estimated signals occur during the synchronization processes within the DT, whereas those in the source signals simply indicate significant changes in the physical system.
Then, from Fig.~\ref{fig:res0} we see that, the proposed CRL algorithm only schedules sensing data transmission for dramatic physical state changes, such that it can follow up with physical system changes with reduced need of data transmission.  
%Fig.~\ref{fig:res0} also shows that the proposed CRL algorithm can capture the waveform of the physical device states, with reduced data transmission. 
The reduction is because the proposed mismatch loss functions can capture variations in physical states, identifying the transmissions of inconspicuous state changes as redundant.
From Fig.~\ref{fig:res0}, we can also observe that
the NRMSE of temperature in~Fig.~\ref{fig:image1} is the lowest as the thermometers have the highest transmission priority weight, 
while the NRMSE of location in~Fig.~\ref{fig:image3} is the highest due to the low transmission priority weight and large RB consumption at positioning sensors.
The difference on estimation precision shown in Fig.~\ref{fig:res0} further validates the proposed CRL algorithm's ability to adapt to different transmission priorities with adjusted device scheduling strategies.

In Fig.~\ref{fig:res1}, we show the changes in the weighted mismatch of all devices with different methods as the number of RBs $M$ increases. 
% In the simulation, the error based method has the fixed RB consumption determined by the physical states, because of the fixed sampling scheme.
Fig.~\ref{fig:res1} shows that as $M$ increases, the weighted mismatch decreases as expected, and the proposed CRL algorithm obtains a lower weighted mismatch than the baselines.
When using 15 RBs, the CRL algorithm can reduce the weighted mismatch by up to 30.58\% and 47.63\%, respectively, compared to the DP method and polling method.
The improvement of the DP method over the polling method  lies in the customized scheduling intervals for diverse devices, which capture the characteristics of physical state changes and thereby reduce the mismatch.
% using the mismatch metric can reduce redundant transmission and perceive channel conditions.
The improvement of the proposed CRL algorithm over the DP method is attributed to the ability to learn the latest varying trends of physical states and channel conditions, thereby adaptively adjusting the scheduling intervals.
Fig.~\ref{fig:res1} also shows that, when $M$ approaches $36$, all the methods exhibit a low mismatch close to $0$ because of the abundant RBs.
However, as $M$ decreases below $10$, the weighted mismatch increases sharply. 
This is due to the fact that the constrained number of available RBs decrease the scheduled data transmission, thereby affecting the mismatch significantly.
%Fig.~\ref{fig:res1} demonstrates that the CRL algorithm can perceive both the variations in physical states and channel conditions, allowing for necessary device scheduling to effectively minimize DT mismatch.

% thus scheduling the devices with necessity to effectively minimize DT mismatch.

\begin{figure}[!t]
    \centering
    \includegraphics[width=1.\linewidth,]{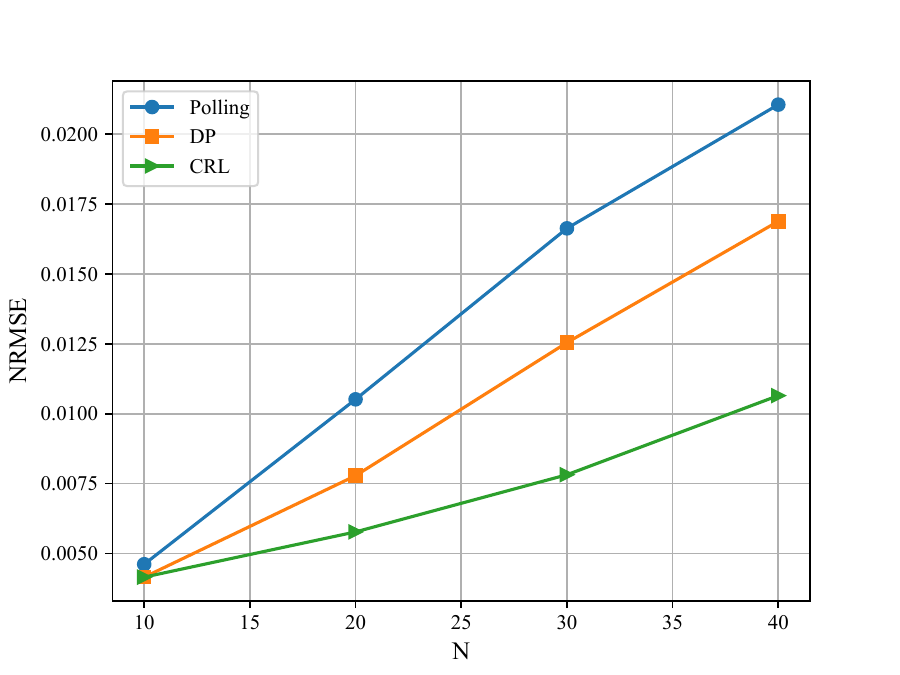}
    \setlength{\abovecaptionskip}{-0.3cm}
    \vspace{-0.2cm}
    \caption{ NRMSE as number of devices $N$ increases, the number of RB constraint is $M=10$.}\label{fig:res2}
    \vspace{0.cm}
\end{figure}

Fig.~\ref{fig:res4} shows how the averaged NRMSE between virtual states and physical states from all devices changes as the number of RBs $M$ increases.
From Fig.~\ref{fig:res4} we observe that, as $M$ increases, the NRMSE decreases because more RBs can support more frequent sensing data transmission, thus reducing the NRMSE of virtual states.
Fig.~\ref{fig:res4} also shows that, when $M$ is close to 36, all methods have a low NRMSE because the RBs are sufficient to schedule all devices in each time slot.
It can also be seen that as $M$ decreases below 10, the NRMSE of all methods increases, with the proposed CRL algorithm exhibiting the slowest increase rate.
This is because that the on-demand synchronization of the frequent and dramatic physical state changes cannot be fully supported by limited number of RBs, leading to an increase of NRMSE at DT. 
The slower NRMSE increase rate of the CRL method stems from the fact that the proposed CRL algorithm continually learns and predicts the physical dynamics, while the polling and DP methods assume regular changes in the physical state.
From Fig.~\ref{fig:res4} we can also see that, to achieve the same NRMSE of $0.02$, the proposed CRL algorithm can reduce the number of consumed RB by up to 55.56\% and 61.54\% compared to the DP method and the polling method, respectively. 
And using the same number of RBs with $M=15$, the proposed CRL algorithm can reduce the NRMSE by up to 55.21\% and 68.42\% compared to the DP method method and polling method, respectively.
The reduction of RB consumptions and NRMSE is because the CRL algorithm learns both the physical state variations and the channel conditions, thus scheduling  the devices with effective data.
Fig.~\ref{fig:res1} and Fig.~\ref{fig:res4} collectively demonstrate that the proposed CRL algorithm offers a more resource-efficient device scheduling solution, particularly in the resource-limited scenarios.

%demonstrates the performance of the proposed CRL algorithm on learning the trade-off relationships between reducing RB consumptions and reducing mismatch for DT synchronization.
% and can reduce the DT mismatch with limited RBs.

% AoII-RL sampling can perceive both the source variation and channel conditions, thus reducing estimation error with semantics.

% \begin{figure}[t]
%     \centering
%     \includegraphics[width=1\linewidth, ]
%     {figs/mis_res/res3_RBcostdist_vs_M.pdf}
%     \setlength{\abovecaptionskip}{-0.1cm}
%     \caption{ Distribution of instantly consumed RBs by the CRL algorithm with different constraints.}\label{fig:res_dist}
%     \vspace{-0.3cm}
% \end{figure}

\begin{figure*}[!t]
    \centering
    \includegraphics[width=1.\linewidth]{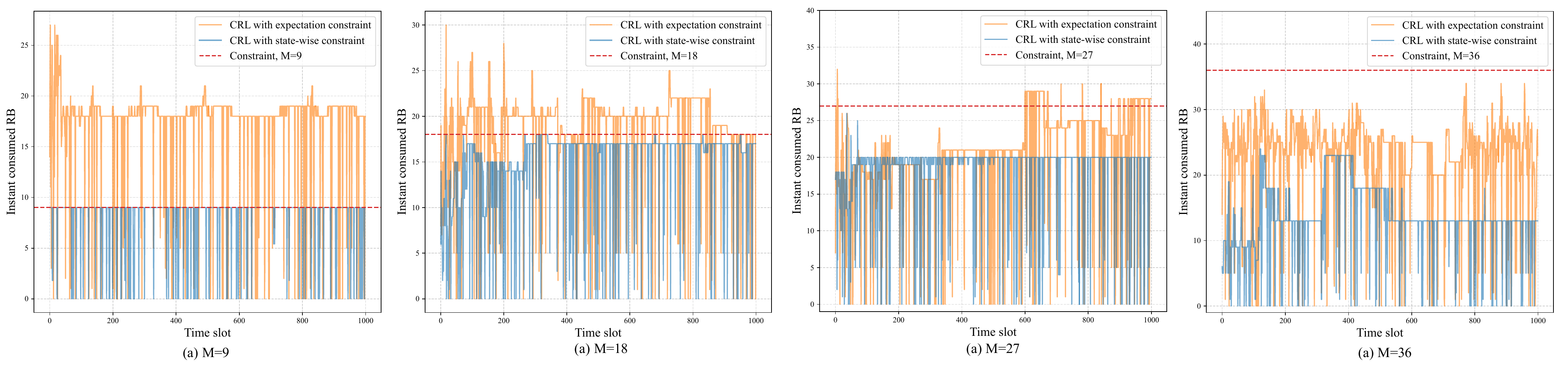}
    \setlength{\abovecaptionskip}{-0.2cm}
    \vspace{-0.2cm}
    \caption{{Instant consumed RBs by the CRL algorithm with different constraints, $M=9,18,27,36$.}}
    \label{fig:4shuzhi}
\end{figure*}  

\begin{figure*}[!ht]
    \centering
    \vspace{-0.3cm}
    \includegraphics[width=1.\linewidth]{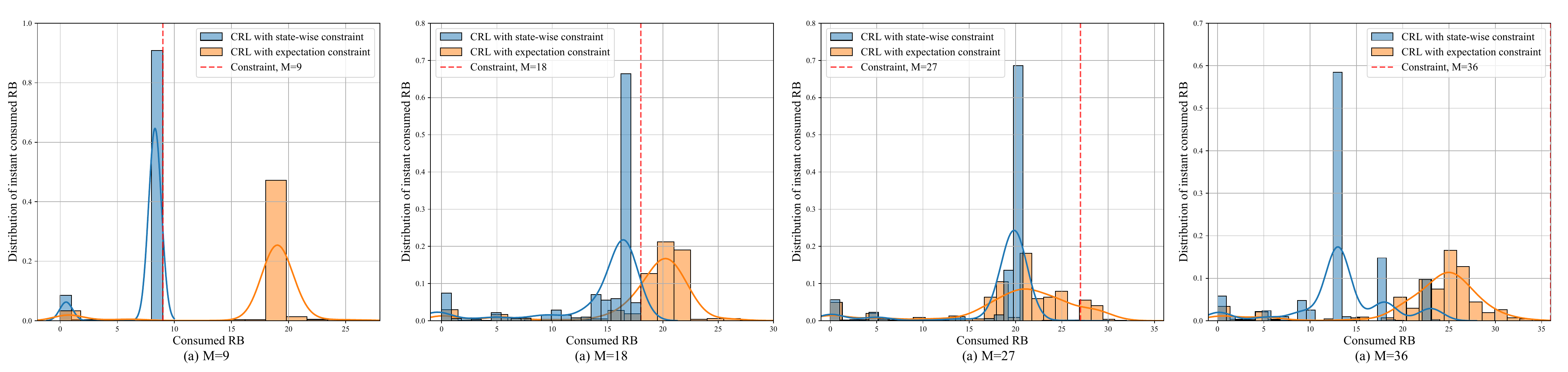}
    \setlength{\abovecaptionskip}{-0.2cm}
    \vspace{-0.2cm}
    \caption{{Distributions of instant consumed RBs, $M=9,18,27,36$. }}
    \label{fig:4fenbu}
\end{figure*}

% \begin{figure*}[!ht]
%     \centering
%     \includegraphics[width=0.99\linewidth]{figs/Major_Revision/4shuzhi.pdf}
%     \setlength{\abovecaptionskip}{0.cm}
%     \caption{{M=36.}}
%     \label{fig:zuheM36}
% \end{figure*}  

Fig.~\ref{fig:res2} shows how the average NRMSE of different methods changes as the number of devices $N$ increases. 
In the simulation, the number of available RB numbers is fixed at $M=10$, and each device consumes $\boldsymbol b_n = 1$ RB per transmission.
From Fig.~\ref{fig:res2} we observe that, when $N=10$, all methods exhibit low NRMSE due to the sufficient RBs, meanwhile, the polling method has a relatively high NRMSE when $N=10$ because of the lack of channel condition perception.
Fig.~\ref{fig:res2} also shows that
as the number of devices $N$ increases, the NRMSE increases nearly linearly due to the increased RB competition among devices.
When $N=20$, the CRL algorithm can reduce the NRMSE by 28.84\% and 45.09\% compared to the DP method and polling method, respectively. 
% From Fig.~\ref{fig:res2}, we can also see that, as $N$ increases, the NRMSE of DP method increases slower than that of polling method, because DP  method can reduce the redundant transmission with the mismatch metric, thus decreasing the mismatch by transmitting more effective data.
% From Fig.~\ref{fig:res2}, we can also see that, as $N$ increases, the NRMSE of DP method increases slower than that of polling method, because DP method extract the physical states varying trends
From Fig.~\ref{fig:res2} we can also see that, the CRL algorithm has the slowest increase rate as the number of devices $N$ increase.
This is because the CRL algorithm learns both the physical states varying trends and channel condition to support high-density device scheduling with limited RBs. 
Fig.~\ref{fig:res2} demonstrates that, compared to the polling and DP methods, the proposed CRL algorithm is more scalable with the enhanced communication resource efficiency in high device density scenarios.

{\color{black}
Fig.~\ref{fig:4shuzhi} shows the values of instant consumed RBs by the proposed CRL algorithm with the expectation constraint and the state-wise constraint when $M=9,18,27,36$.
This simulation is conducted to demonstrate the constraint satisfaction performance of solving the state-wise constrained problem.
From Fig.~\ref{fig:4shuzhi}-(a) and (b) we see that, under resource-scarce conditions ($M=9,$ and~$18$), the expectation constraint method will tentatively exceed the constraint value $M$ with a significant probability.
In contrast, the proposed method, which learns CRL with a state-wise constraint, can significantly guarantee that the instant consumed RBs under the RB capacity constraint.
This is because that the state-wise constraint problem formulation imposes stricter limitations on the CRL algorithm, ensuring that RB usage conforms better to practical conditions.
Fig.~\ref{fig:4shuzhi}-(c) and (d) show that under resource-sufficient conditions ($M=27,$ and~$36$), the state-wise constraint method uses fewer resources compared to the expectation constraint method, ensuring the RB consumption satisfies the constraint.
This is because the cost definition in the proposed CRL method includes a penalty for exceeding the RB capacity, causing the CRL to  reduce RB usage thus decreasing the probability of exceeding the capacity constraint.
Fig.~\ref{fig:4shuzhi} demonstrates that the proposed method effectively ensures the satisfaction of the RB constraints.

Fig.~\ref{fig:4fenbu} shows the distributions of instant consumed RBs in Fig.~\ref{fig:4shuzhi}, for $M=9,18,27,36$. 
From Fig.~\ref{fig:4fenbu} we can observe that, the state-wise constraint method results in distributions with smaller means and variances, compared to the expectation constraint method.
% From Fig.~\ref{fig:4fenbu}, we can observe that the probability distribution of the instant consumed RBs by the state-wise constraint method has smaller mean and variance compared to the expectation constraint method.
This phenomenon is attributed to the state-wise constraint compelling the CRL method to use fewer RBs. 
Furthermore, the IRM mechanism stabilizes the policy output of the CRL method, thus enabling more effective DT synchronization optimization.
}

% by the proposed CRL algorithm under different constraints $M$. 

% From Fig.~\ref{fig:res_dist}, we observe that, the instantaneous sampling rate increases as~$\delta$ increases. 
% This is because a higher sampling rate can extract more semantic information from sensing data and reduce its AoII.
% From Fig.~\ref{fig:res_dist}, we can also see that, under different sampling rate constraints~$\delta$, the instantaneous sampling rates of AoII-RL sampling  remain within the constraint with a high probability. 
% This phenomenon is due to the fact that, the proposed SAC-based RL agent can efficiently learn the trade-off relationship between AoII and sampling rate, thus minimizing long-term AoII within sampling rate constraint.
% Fig.~\ref{fig:res_dist} shows the statistics of instantaneous sampling rates  for  AoII-RL sampling with different sampling rate constraint~$\delta$. 
% From Fig.~\ref{fig:res_dist}, we observe that, the instantaneous sampling rate increases as~$\delta$ increases. 
% This is because a higher sampling rate can extract more semantic information from sensing data and reduce its AoII.
% From Fig.~\ref{fig:res_dist}, we can also see that, under different sampling rate constraints~$\delta$, the instantaneous sampling rates of AoII-RL sampling  remain within the constraint with a high probability. 
% This phenomenon is due to the fact that, the proposed SAC-based RL agent can efficiently learn the trade-off relationship between AoII and sampling rate, thus minimizing long-term AoII within sampling rate constraint.

\section{Conclusion}
In this paper, we have developed a CRL based adaptive resource allocation scheme to optimize DT synchronization over dynamic wireless networks.
To maintain DT synchronization with limited wireless resources, we proposed to selectively schedule devices for data transmission, meanwhile optimize the allocation of RBs over the wireless transmission links.
% schedules devices to collect effective data, which can revise the DT mismatch.
A CMDP problem has been formulated to capture the goal of minimizing long-term mismatch between the physical and virtual systems over device scheduling decisions.
% Based on the defined mismatch model according to the physical state character, we have formulated a CMDP problem that minimizes the long-term weighted mismatch. 
To solve the problem, we first transformed the original problem into a dual problem to refine the impacts of RB constraints on device scheduling strategies.
Then we proposed a CRL algorithm to solve the dual problem.
The CRL algorithm consists of an MTR buffer that stabilizes the SAC policy learning processes across historical experiences, which achieves quick adaptation of device scheduling solutions to the dynamics in physical object states and network capacity.
% then a CRL resource allocation policy consisting of an MTR buffer and SAC agent was proposed to guarantee that the effective data can be collected in time thus reducing the mismatch in the long-term DT mapping.
Simulation results have demonstrated that the proposed CRL algorithm is resilient to the changes in wireless network capacity, and shows higher wireless resource efficiency, even in resource-limited, high device density scenarios. 
The proposed CRL algorithm can reduce NRMSE of the estimated virtual states by up to 55.2\%, using the same number of RBs as traditional methods.

\begin{appendix}
\subsection{Proof of Theorem 1}

% Construct an alternative constrained optimization problem here, whose formulation is exactly
% corresponding to the optimization problem $\mathbb{P}2$:

% \boldsymbol Z

First, denote $v(\boldsymbol{Z}) = \frac{1}{N} \boldsymbol{w}^{\top} \boldsymbol{Z}$, the problem $\mathbb{P}2$ can be reformulated by

\begin{equation} \label{equ:A-PL}
\begin{aligned}
\min_\pi & ~\mathbb{E}_{\boldsymbol Z \sim d_0(\boldsymbol Z)} v(\boldsymbol Z) \\
\text { s.t. } &~ d_0(\boldsymbol Z)\left[V_c(\boldsymbol Z)-M\right] \leq 0, \forall \boldsymbol Z \in \mathcal{S}_F,
\end{aligned}
\end{equation}
where $d_0(\boldsymbol Z)$ is the practical sampled state distribution and $\mathcal{S}_F$ is the feasible region of state.

We consider the constraint of problem~(\ref{equ:A-PL}). The distribution of initial state $d_0(\boldsymbol Z)$  has the property of
$d_0(\boldsymbol Z) \geq 0$.
While the initial feasible state set has the property
\begin{equation}
\mathcal{I}=\left\{\boldsymbol Z \mid d_0(\boldsymbol Z)>0\right\}.
\end{equation}
Because that for those states not in the possible initial state set, which is  $ \boldsymbol Z \notin \mathcal{I}$, we have $d_0(\boldsymbol Z) = 0$, according to the definition of possible initial state set.
Therefore, the constraint in~(\ref{equ:A-PL}) can be reformulated to
\begin{equation}
V_c(\boldsymbol Z)-M \leq 0, \forall \boldsymbol Z \in \mathcal{I} \cap \mathcal{S}_F ,
% {\mathcal{S}_F},
\end{equation}
which is equivalent to the constraint in $\mathbb{P}2$
% According to the definition of possible initial state set, for those $z$ with $d_0(z) = 0$, $z \notin \mathcal{I}$. For
% those $z$ with $d_0(z) =  0$, that is, $z \in \mathcal{I}$ the constraints in problem (\ref{equ:A-PL}) is equivalent to  $[V_c(\boldsymbol{Z})-M] \leq 0$. Therefore, the constraint can be reformulated to
\begin{equation}
V_c(\boldsymbol Z)-M \leq 0, \forall \boldsymbol Z \in \mathcal{I}_{\mathcal{F}}.
\end{equation}
% Hence, the problem in~(\ref{equ:A-PL}) is 

Since it is infeasible to find a policy which enforces infeasible states to satisfy the constraint,  consider that the state distribution  $d_0(\boldsymbol Z )$ can be sampled from the feasible region $\mathcal{S}_F$, and an alternative Lagrangian function form of $\mathcal{L}_\text{stw}(\boldsymbol{u}, \lambda)$ in equation~(12)
% \ref{equ:stw}
 % with policy $\pi(\theta)$ 
 can be given by 
\begin{equation}
\begin{aligned}
& \mathcal{L}_{\text{stw}}'(\boldsymbol{u}, \lambda)\\ 
& =\mathbb{E}_{\boldsymbol Z  \sim d_0(\boldsymbol Z )}\left\{v(\boldsymbol Z )+\lambda(\boldsymbol Z )\left[V_c(\boldsymbol Z )-M\right]\right\} \\
& =\mathbb{E}_{\boldsymbol Z  \sim d_0(\boldsymbol Z )}\left\{v(\boldsymbol Z )\right\}+\sum_{\boldsymbol Z  \in \mathcal{S}_F} d_0(\boldsymbol Z ) \lambda(\boldsymbol Z )\left[V_c(\boldsymbol Z )-M\right] \\
& =\mathbb{E}_{\boldsymbol Z  \sim d_0(\boldsymbol Z )}\left\{v(\boldsymbol Z )\right\}+\sum_{\boldsymbol Z  \in \mathcal{S}_F} \lambda(\boldsymbol Z )\left\{d_0(\boldsymbol Z )\left[V_c(\boldsymbol Z )-M\right]\right\} , 
\end{aligned}
\end{equation}
which is exactly the Lagrangian function of problem~(\ref{equ:A-PL}).
Hence, $\mathcal{L}_{\text{o-stw}}(\boldsymbol{u}, \lambda)$ is equivalent to  $\mathcal{L}_\text{stw}(\boldsymbol{u}, \lambda)$.
$\hfill \blacksquare$
% \qed

\end{appendix}

\bibliographystyle{IEEEtran}
\bibliography{IEEEabrv,ref}

\end{document}